\documentclass[a4,11pt]{article}
\pdfoutput=1 
\usepackage{jheppub}
\usepackage{graphicx}
\usepackage{color, graphics}
\usepackage{amsmath,amssymb,amsfonts}
\usepackage{acronym}
\usepackage{mathrsfs}
 \usepackage{hyperref}
 \usepackage{units}
\usepackage{enumerate}   
\usepackage{etoolbox}

\usepackage{tikz}
\usetikzlibrary{quotes,arrows.meta,calc,patterns,angles}
\usetikzlibrary{%
    decorations.pathreplacing,%
    decorations.pathmorphing%
}
\usetikzlibrary{decorations.markings}
\usetikzlibrary{positioning,arrows,shapes} 
\tikzset{photon/.style={decorate, decoration={snake}, draw=black},
fermion/.style={thick,draw=blue, postaction={decorate},
    decoration={markings,mark=at position .5 with {\arrow[blue]{triangle 45}}}},
gluon/.style={decorate, draw=black,
    decoration={coil,aspect=0}},
scalar/.style={thick,dashed,draw=blue, postaction={decorate}},
higgs/.style={thick,loosely dashed,draw=blue, postaction={decorate}}}    

\def\beq{\begin{equation}\begin{aligned}}
\def\eeq{\end{aligned}\end{equation}}

\begin{document}

\title{Improved Constraints on Dark Matter Annihilations Around Primordial Black Holes}
\author[1]{Prolay Chanda,}
\author[2]{Jakub Scholtz,}
\author[1,3,4]{and James Unwin}
\affiliation[1]{Department of Physics,  University of Illinois at Chicago, Chicago, IL 60607, USA}
\affiliation[2]{Dipartimento di Fisica, Universit\`a di Torino, via P. Giuria 1, 10125, Torino, Italy}
\affiliation[3]{Physics Division, Lawrence Berkeley National Laboratory, Berkeley, CA 94720, USA}
\affiliation[4]{Berkeley Center for Theoretical Physics, University of California, Berkeley, CA 94720, USA}

\abstract{Cosmology may give rise to appreciable populations of both particle dark matter and primordial black holes (PBH) with the combined mass density providing the observationally inferred value $\Omega_{\rm DM}\approx0.26$. Early studies  highlighted that scenarios with both particle dark matter and PBH are strongly excluded by $\gamma$-ray limits for particle dark matter with a velocity independent thermal cross section  $\langle\sigma v\rangle\sim3\times10^{-26}{\rm cm}^3/{\rm s}$, as is the case for classic WIMP dark matter. Here we examine the limits from diffuse $\gamma$-rays on velocity-dependent, including annihilations which are $p$-wave with $\langle\sigma v \rangle\propto v^2$ or $d$-wave  $\langle\sigma v \rangle\propto v^4$, which we find to be considerably less constraining. This work also utilyses a refined treatment of the PBH dark matter density profile. Importantly, we highlight that even if the freeze-out process is $p$-wave it is typical for (loop/phase-space) suppressed $s$-wave processes to actually provide the leading contributions to the experimentally constrained $\gamma$-ray flux from the PBH halo. 
}

\maketitle

\section{Introduction}

Many scenarios of early universe cosmology lead to the formation of primordial black holes (PBH)  \cite{Zeldovich:1967lct, Hawking:1971ei,Carr:1974nx,Carr:1975}. Provided that the PBH have mass $M_{\bullet}\gtrsim10^{11}$ kg, they will survive to the present day, while PBH below this mass bound evaporate in the early universe due to the emission of Hawking radiation \cite{Hawking:1974rv}. Assuming a monochromatic mass spectrum, PBH of mass  $10^{-16} M_\odot\lesssim M_{\bullet}\lesssim 10^{-10} M_\odot$ can exist as astrophysical objects and account for all of the dark matter without being constrained by experiments \cite{Carr:2020xqk,Carr:2020gox}. Furthermore, in the range $10^{-10} M_\odot\lesssim M_{\bullet}\lesssim 10^{-7} M_\odot$, and a small window between $M_{\bullet}\sim1-10 M_\odot$, PBH  can account for up to 1-10\% of the inferred dark matter relic density $\Omega_{\rm DM}\approx0.26$ before running into experimental limits.  Thus, from an observational perspective, it is quite plausible that the universe could host both particle dark matter along (pDM) with a non-negligible abundance of PBH. Notably, in this ``mixed PBH-pDM'' scenario one anticipates that the particle dark matter will form dense halos around the PBH. As a result, these PBH halos can lead to enhanced dark matter self annihilations, leading to detectable signals.

The prospect of PBH and dark matter coexisting in our universe has previously be explored in a number of papers (e.g.~\cite{Dokuchaev:2001ekr,Mack:2006gz,Ricotti:2009bs,Ricotti:2007jk,Eroshenko:2016yve,Lacki:2010zf,Kadota:2021jhg,Adamek:2019gns,Carr:2020mqm,Scholtz:2019csj,Cai:2020fnq,Tashiro:2021xnj,Boudaud:2021irr,Kadota:2022cij,Gines:2022qzy})  in the context of WIMP dark matter with an $s$-wave thermal annihilation cross section $\langle\sigma v\rangle\sim3\times10^{-26}{\rm cm}^3/{\rm s}$. These earlier papers found a striking result, namely that for the $s$-wave WIMP case $\gamma$-ray bounds imply that the abundance of either particle dark matter or PBH must be vanishingly small in order to avoid exclusion limits. In this work, we extend these previous studies by using a more careful analysis of the PBH halo profile and, more importantly, ascertaining the limits on velocity dependent $p$ and $d$ wave thermal dark matter.
       
       To put the present work into context, let us first discuss the existing literature. The first papers on dark matter accretion by PBH \cite{Dokuchaev:2001ekr, Ricotti:2007jk, Ricotti:2009bs,Mack:2006gz} used the simplest assumption of radial infall of dark matter particles, leading to a spike-like dark matter density profile around the PBH with $\rho(r)\sim r^{-9/4}$, according to the famous result by Bertschinger \cite{Bertschinger1985}. Eroshenko \cite{Eroshenko:2016yve} subsequently presented a more detailed study of dark matter halo formation around a PBH taking into account the orbital motion of the particles, although the results were restricted to inert dark matter with a mass around 70 GeV. The most sophisticated study of the dark matter halos around PBH is from a recent paper of Boudaud {\em et al.}~\cite{Boudaud:2021irr} who provide both a numerical study and an analytical approximation for the density profiles. This work explains how earlier calculations are incomplete, presenting the state-of-the-art understanding of dark matter density profiles, and thus it is this framework that we utilise as our starting point. 
       
As mentioned, if dark matter can annihilate to Standard Model states, as is assumed in the classic WIMP picture, then one would expect observable signals from annihilations within the dark matter halo of the PBH. Lacki \& Beacom \cite{Lacki:2010zf} were the first to explore these implications, declaring ``Primordial Black Holes as Dark Matter: Almost All or Almost Nothing'' since their findings placed extremely strong limits on this mixed PBH-dark matter scenario for WIMP dark matter with certain assumptions.  These limits have been re-examined for simple WIMP dark matter in subsequent papers, for instance Adamek {\em et al.}~\cite{Adamek:2019gns},  who repurposed the bounds on decaying dark matter, and analyses using constraints on the extragalactic background photon fluxes \cite{Carr:2020mqm,Gines:2022qzy}, among others \cite{Bertone:2019vsk,Scott:2009tu,Boucenna:2017ghj}. 

Notably, the bulk of the existing literature focuses entirely on bounds on WIMP dark matter with an $s$-wave annihilation route.  In this paper we seek to extend this analysis beyond $s$-wave dark matter to models with velocity dependent dark matter annihilation cross sections. We present bounds on PBH dark matter halos in the context of the careful halo density profile calculations laid out in \cite{Boudaud:2021irr}.
We note that two recent papers    \cite{Gines:2022qzy,Kadota:2021jhg}  considered $p$-wave dark matter, however  here we present more sophisticated and detailed treatments of this class of models, including the mixed $s$-wave/$p$-wave case. We highlight that in all realistic models of $p$-wave dark matter, one expects a non-zero $s$-wave contribution (which in many cases is suppressed only by a few orders of magnitude). We show in Section \ref{S6} that including the $s$-wave contribution is critical to reliably calculating the $p$-wave limits. Indeed, it might be argued that this invalidates the strong statements made regarding the large amelioration of limits on $p$-wave dark matter derived in the earlier works  \cite{Gines:2022qzy,Kadota:2021jhg}.

This work is structured as follows: in Section \ref{S2} we outline how the dark matter density profile around the PBH is calculated, we follow, in particular, the recent improved treatment of  Boudaud {\em et al.}~\cite{Boudaud:2021irr}.  Subsequently, in Section \ref{S4} we continue to examine the important impacts of dark matter interactions on the halo profile which goes beyond the analysis of \cite{Boudaud:2021irr}.  We compare the resulting profiles to other treatments in the literature finding general consistency. In Section \ref{S5} we derive the limits on dark matter annihilation in the PBH halo for thermal dark matter with annihilation cross section which are either $s$, $p$, or $d$ wave, i.e.~velocity independent or suppressed by factors of $v^2$ or $v^4$. We interpret these bounds on dark matter annihilation  in terms of a limit on $f_{\rm PBH}$, the fraction abundance of PBH contributions to $\Omega_{\rm DM}$.  In Section \ref{S6} we re-examine the $p$-wave scenario with the more realistic assumption that the $s$-wave component is non-zero, but suppressed, and show that this can have a significant impact on the results. We present  some concluding remarks in Section \ref{S7}. The Appendices include discussions on the potential impact of dark matter self interactions, an example model of $d$-wave annihilating dark matter, and a comparison to earlier studies.

\section{Dark Matter Around Primordial Black Holes}
\label{S2}

In this section we discuss the formation of dark matter halos around PBH, outlining first the initial distribution of matter around the PBH and then discussing the late time dark matter density profiles $\rho(r)$ neglecting dark matter interactions. Subsequently, in Section \ref{S4} we discuss the impact of dark matter annihilation to the PBH halo profile.

\subsection{PBH Formation}

A primordial black hole is formed as a density perturbation enters the horizon, and thus the mass of the black hole will be close to the mass of the matter inside that region, $M_{\rm horizon}$, i.e.~$M_{\rm horizon} \sim M_{\bullet}$ \cite{Khoplov:1980}. Assuming the Newtonian limit to the FLRW universe, the kinematics of a shell of particles at a distance $r$ from the PBH is given by (see e.g.~\cite{Adamek:2019gns})
\beq\label{eq:DMmotionPBH1}
\ddot{r} = -\frac{GM_{\bullet}}{r^{2}}+\frac{\ddot{a}}{a}r.
\eeq
The two terms on the RHS correspond to the gravitational field of the PBH, and the decelerating background expansion, respectively. The time at which  these two terms become comparable in magnitude marks a special point, commonly called the ``turn-around'' time $t_{\rm ta}$, at which point in time the mass shell decouples from the Hubble flow and re-collapses towards the black hole. Accordingly, one can evaluate the radius of the shell in eq.~(\ref{eq:DMmotionPBH1}) at this time, defining the turn-around radius $r_{\rm ta}\equiv r(t_{\rm ta})$, given by
\beq\label{eq:turnaroundRadius2}
r_{\rm ta} = \left(\frac{8GM_{\bullet}t_{\rm ta}^{2}}{1+3\omega}\right)^{1/3},
\eeq
where $\omega$ is the FLRW equation of state (with $\omega=1/3$ for a radiation dominated universe).\footnote{Analysis in \cite{Adamek:2019gns} suggests that $r_{\rm ta} = \left(2\eta_{\rm ta}GM_{\bullet}t_{\rm ta}^{2}\right)^{1/3}$ provide a slightly better fit to numerical results (where $\eta_{\rm ta} =  1.074$), note that the prefactor of this form differs slightly to that of eq.~(\ref{eq:turnaroundRadius2}).}
Taking this relationship, along with the fact that at matter-radiation equality the total dark matter mass is equal to the energy in radiation, it follows that $r_{\rm \rm  ta}$ can be understood as the radius of influence of the black hole.  The radius of influence evaluated at matter-radiation equality (with $t=t_{\rm eq}$) can be defined as the size of the halo at this time:
\beq\label{eq:rtildeeq}
  \tilde{r}_{\rm eq}\equiv \frac{r_{\rm ta}(t_{\rm eq})}{r_{\rm Sch}}  =\left(\frac{\eta_{\rm ta}c^{2}}{4G^{2}}\right)^{1/3}M_{\bullet}^{-2/3}t_{\rm eq}^{2/3},
\eeq
where $r_{\rm Sch} = 2GM_{\bullet}/c^{2}$, is the Schwarzschild radius.

The formation time for a primordial black hole, $t_{\rm form}$, is related to the PBH mass via $t_{\rm form} \sim GM_{\bullet}$. Moreover, during radiation domination time scales with temperature as follows
$  t = (45/16G\pi^{3}g_{\rm eff})^{1/2} T^{-2}$, where $g_{\rm eff}$ is the number of relativistic degrees of freedom at temperature $T$.
   Thus, if a primordial black hole is formed in a radiation dominated universe at temperature $T$, then the mass of the black hole is  $M_{\bullet} \sim \sqrt{5/(48\pi^{3}G^{3}g_{\rm eff})}~T^{-2}$, and it follows that heavier PBH form at later times. 

\subsection{Kinetic Decoupling of Dark Matter}
   
    Prior to kinetic decoupling (at $t=t_{\rm kd}$) the dark matter particles cannot be  significantly captured by the PBHs as they are in equilibrium with the background plasma. If the dark matter is kinetically decoupled prior to $t=t_{\rm ta}$, then the ball of dark matter around the PBH becomes gravitationally bound at this point.  Thus, for PBHs formed before kinetic decoupling, a  dark matter halo of constant density  equal to the background dark matter density at kinetic decoupling  will form around the black hole: 
     $\rho(t_{\rm kd})=\rho_c\Omega_{\rm pDM}(m_\chi/x_{\rm kd}T_0)$,
where $\rho_c$ and $T_0$ are the critical density and temperature today, $x_{\rm kd}$ is the inverse scaled temperature ($x=m/T$)  at $T(t_{\rm kd})$, and $\Omega_{\rm pDM}$ is abundance of particle dark matter today. Throughout we will use $\chi$ to indicates dark matter particle with mass $m_\chi$.

  We can parameterise  the kinetic decoupling of dark matter particles such that this occurs when the  temperature of the thermal bath is $T_{\rm kd} =  \alpha_{\rm kd} m_{\chi}$, where $\alpha_{\rm kd}$ is a constant which parameterises the point of decoupling. In this case one can define a characteristic PBH mass
    \beq
   M_{c} = \left(\frac{5}{48\pi^{3}G^{3}g_{\rm eff}(T_{\rm kd})\alpha_{\rm kd}^{4}}\right)^{1/2}\frac{1}{m_{\chi}^2},
   \eeq
    such that black holes with masses $M_{\bullet}>M_{c}$ form after the dark matter particles are already decoupled from the thermal bath. Conversely, PBH  with masses $M_{\bullet}<M_{c}$ form before the dark matter is kinetically decoupled and, therefore,  accretion of radiation affects the initial dark matter halo until the point of dark matter decoupling \cite{Eroshenko:2016yve}.

For PBH which form much later than $t_{\rm kd}$ the background dark matter density evolves as $\rho_{\rm DM}\propto T^{3}$, where $T$ is the temperature of the background. We can define a length scale inside which the dark matter particles become kinetically decoupled by calculating $r_{\rm ta}(t_{\rm kd})$ (the turnaround radius  evaluated at kinetic decoupling)
 \beq\label{eq:rtildekd}
   \tilde{r}_{\rm kd}\equiv \frac{r_{\rm ta}(t_{\rm kd})}{r_{\rm Sch}} = \left(\frac{\eta_{\rm ta}c^{2}}{4}\right)^{1/3}G^{-2/3}M_{\bullet}^{-2/3}t_{\rm kd}^{2/3}.
 \eeq
    The density profile of the dark matter halos around the heavier black holes which form after the kinetic decoupling should evolve as $\rho(r)\propto t^{-3/2}$; then using eq.~\eqref{eq:turnaroundRadius2}, one finds that the dark matter density prior to collapse scales as: $\rho(r)\propto r^{-9/4}$ \cite{Bertschinger1985}.

Putting this together it follows that the dark matter density before collapsing into the halo, after kinetic decoupling, can be expressed as \cite{Boudaud:2021irr}
\beq\label{eq:PreCollapse-density}
\rho_{i}(r_{i}) = \left\lbrace
\begin{array}{ll}
~\rho_{i}^{\rm kd}&\qquad r_{i}\leq r_{\rm kd}\\[15pt]
~\rho_{i}^{\rm kd}\left(r_{i}/r_{\rm kd}\right)^{-9/4}&\qquad r_{\rm kd}\leq r_{i}\leq r_{\rm eq}
\end{array}
\right..
\eeq
we use the subscript `$i$' here to indicate that these are `initial' quantities, prior to gravitational collapse.
Moreover, following kinetic decoupling, the dark matter halo density evolves, and grows beyond $r_{\rm kd}$, via secondary accretion leading to the formation of a density spike. As we will derive below, the late time density profile (neglecting dark matter interactions) can involve multiple regions, which scale according to different power-laws involving the radial distance $r$ from the centre of mass.

\subsection{PBH Halo Profiles Neglecting Dark Matter Interactions}

To obtain the detailed late time density profile of the PBH dark matter halo one also needs to consider the dark matter particle velocity distribution. Notably, the dark matter particles on the lower velocity tail of the velocity distribution are going to stay captured, while the higher velocity ones escape the PBH. Before dark matter kinetic decoupling occurs, the dark matter particles are in thermal contact with the background plasma, and thus follow a Maxwell-Boltzmann distribution. Following kinetic decoupling this thermal contact breaks and the dark matter momenta redshift as $1/a$.  

The one-dimensional velocity dispersion can be written  $\sigma = x^{-1/2}$  and one can express the pre-collapse velocity distribution in a piecewise form (similar to the pre-collapse density), in terms of the velocity dispersion at kinetic decoupling $\sigma_{\rm kd} = x_{\rm kd}^{-1/2}$, as follows \cite{Boudaud:2021irr}
\beq\label{eq:PreCollapse-velocity}
\sigma_{i}(r_{i}) = \left\lbrace
\begin{array}{ll}
\sigma_{i}^{\rm kd}& r_{i}\leq r_{\rm kd}\\[15pt]
\sigma_{i}^{\rm kd}\left(r_{i}/r_{\rm kd}\right)^{-3/2}&r_{\rm kd}\leq r_{i}\leq r_{\rm eq}
\end{array}
\right.~.
\eeq
It is important to incorporate the orbital motion of the dark matter particles as studied by \cite{Eroshenko:2016yve,Boudaud:2021irr}, and, also, numerically by \cite{Adamek:2019gns}. 
The late time dark matter density profile around the PBH $\rho(r) $ is obtained by taking the initial distribution given in terms of the pre-collapse radial distance $r_{i}$ and velocity $v_{i}$, and integrating it taking into account the orbital motion of the dark matter particles.
These derivations make the simplifying assumption that the dark matter particles are on circular orbits.

Expressing the radial distance in units of Schwarzschild radius $\tilde r=r/r_{\rm Sch}$, and defining the pre-collapse dark matter velocity relative to the speed of light  $\beta_{i}\equiv v_i/c$, the density profile can be computed by evaluating the following expression~\cite{Boudaud:2021irr}
\beq\label{eq:DensityMiniSpike}
\rho(\tilde{r}) = \frac{8}{\tilde{r}} \int_{0}^{\infty}d\beta_{i}\beta_{i}\int_{0}^{\infty}d\tilde{r}_{i}\tilde{r}_{i}\rho_{i}(\tilde{r}_{i})f(\beta_{i},\tilde{r}_{i})\left(\frac{1}{\tilde{r}_{i}} - \beta_{i}^{2}\right)^{3/2}\int_{ \sqrt{\mathcal{Y}_{m}}\Theta(\mathcal{Y}_m-0)}^{1}\frac{dy}{\sqrt{y^{2}-\mathcal{Y}_{m}}},
\eeq
where $f(\beta_{i},\tilde{r}_{i})$ is the fraction of dark matter particles with velocities between $\beta_i$ and $\beta_i+{\rm d}\beta_i$ given by
\beq
4\pi \beta_i^2{\rm d}\beta_i f(\beta_{i},\tilde{r}_{i})=\frac{4\pi \beta_i^2}{(2\pi \sigma_i^2)^{3/2}}\exp\left(-\frac{\beta_i^2}{2 \sigma_i^2}\right){\rm d}\beta_i 
\eeq
and defining
\beq
\mathcal{Y}_{m} = 1+\frac{\tilde{r}^{2}}{\tilde{r}_{i}^{2}}\left[\frac{1}{\beta_{i}^{2}}\left(\frac{1}{\tilde{r}_{i}} - \frac{1}{\tilde{r}}\right)-1\right].
\eeq
Note that the integral with respect to radius accounts for the pre-collapse density profile $\rho_i(r_i)$, while the integral over $y$ incorporates the integration over the angular variable of the dark matter orbits.

While one can obtain the dark matter density profile by numerically solving eq.~\eqref{eq:DensityMiniSpike}, here we shall employ the approximate analytic solutions as derived in \cite{Boudaud:2021irr}. A simpler form for eq.~\eqref{eq:DensityMiniSpike} can be found by replacing the variables $\beta_{i}, \tilde{r}_{i}$ by normalized variables 
\beq
\mathcal{R} &= \tilde{r}/\tilde{r}_{i},\\
u &= \beta_{i}^{2}\tilde{r}_{i},\\
 \bar{u}_{i} &= \sigma_{i}^{2}\tilde{r}_{i}~.
\eeq
The quantity $u$ can be understood as a ratio between the kinetic energy and the potential energy of a particle. As a result if $u>1$ a particle is unlikely to remain bound to the PBH, and conversely. In terms of these variable the density profile takes a form of:
\beq\label{eq:DensityMiniSpike1}
\rho(\tilde{r}) = \sqrt{\frac{2}{\pi^{3}}}\iint {\rm d}\mathcal{R}{\rm d}u ~\rho_{i}(\mathcal{R}\tilde{r})\frac{\exp\left(-u/2\bar{u}_{i}\right)}{\bar{u}_{i}^{3/2}}(1-u)^{3/2}\mathcal{F}(\mathcal{Y}_{m}),
\eeq
where $\mathcal{F}(\mathcal{Y}_{m})$ is the contribution from the angular integration over $y$. 

As the universe evolves, both the dark matter density $\rho_{i}$ and the velocity dispersion $\sigma_{i}$ decreases as functions of the cosmological scale factor $a$. Following kinetic decoupling of the dark matter particles, the dark matter energy density scales as $\rho_{\chi} \sim a^{-3}$, whereas, the particle dark matter momentum scales as $p_{\chi} \sim a^{-1}$, thus it follows that $\rho_{i}/\sigma_{i}^{3}\approx \rho_{i,{\rm kd}}/\sigma_{i,{\rm kd}}^{3} $ which can be used to simplify eq.~\eqref{eq:DensityMiniSpike1}. Moreover, since the dispersion relationship goes like $\exp(-\beta^2/\sigma^2) \sim \exp(-u/\bar{u})$, we may approximate the exponential in eq.~\eqref{eq:DensityMiniSpike1} with a Heaviside theta function $\Theta(\bar{u}_{i}-u)$, for further simplification. Note that the step function must be dressed with a numerical matching factor in order to agree with the results obtained from the Gaussian distribution. Although in the majority of the studies this factor is assumed to be $\mathcal{O}(1)$,  use of an exact  matching factor is needed for more precise calculation, see e.g.~\cite{Boudaud:2021irr}.  Making use of these simplifications, eq.~\eqref{eq:DensityMiniSpike1} reduces to the following form
\beq\label{simp}
\rho(\tilde{r})\approx \sqrt{\frac{2}{\pi^{3}}}\frac{\rho_{i,{\rm kd}}}{\sigma_{\rm kd}^{3}}\tilde{r}^{-3/2}\iint {\rm d}\mathcal{R} {\rm d}u\lbrace\mathcal{R}(1-u)\rbrace^{3/2}\Theta\left(\bar{u}_{i}-u\right)\mathcal{F}(\mathcal{Y}_{m}).
\eeq
When evaluating the integration in eq.~\eqref{simp}, it is helpful to define the width $\bar{u}\equiv \sigma_{i}^{2}\tilde{r_{i}}$, and note that at kinetic decoupling     %
   \beq
    \bar{u}_{\rm kd} = \sigma_{\rm kd}^{2}\tilde{r}_{\rm kd} = 
    \left(\frac{45\eta_{\rm ta}c^{2}}{64\pi^{3}g_{\rm eff}(T_{\rm kd})  G^3} 
    \frac{1}{m_{\chi}^{3}T_{\rm kd}M_{\bullet}^{2}}\right)^{1/3}, 
    \eeq
where we have used $\sigma_{\rm kd} = \sqrt{T_{\rm kd}/m_{\chi}}$.
Similarly, we can calculate the width at the matter-radiation equilibrium:
   \beq
    \bar{u}_{\rm eq} = \sigma_{\rm eq}^{2}\tilde{r}_{\rm eq} =\sqrt{\frac{45}{16\pi^{3}}}\left(\frac{\eta_{\rm ta}c^{2}}{4G^{7/2}}\right)^{1/3}\frac{(g_{\rm eff}(T_{\rm eq}))^{1/6}}{(g_{\rm eff}(T_{\rm kd}))^{2/3}}t_{\rm eq}^{-1/3}\frac{M_{\bullet}^{-2/3}}{m_{\chi}T_{\rm kd}}, 
    \eeq
where $\sigma_{\rm eq}\equiv \sigma(r_{\rm eq})$  is calculated using eq.~\eqref{eq:PreCollapse-velocity} by evaluating with $r=r_{\rm eq}$.

As the PBH mass varies, there are three different regimes in which the  analytical form of the halo profile around the PBH has distinctly different forms, defined by 
   \begin{itemize}
   \item  ~~$\bar{u}_{\rm eq},\bar{u}_{\rm kd}>1$,
   \item  ~~$\bar{u}_{\rm eq}\leq 1\leq \bar{u}_{\rm kd}$,
   \item  ~~$\bar{u}_{\rm eq},\bar{u}_{\rm kd}<1$. 
      \end{itemize}
      Transitions between these regimes occur at mass scales $M_{1}$, and, $M_{2}$
   given by
    \beq\label{M1M2}
M_{1}& =\frac{1}{16}\left(\frac{45}{\pi^{3}g_{\rm eff}(T_{\rm eq})}\right)^{3/4}\sqrt{\frac{\eta_{\rm ta}}{m_{\chi}^{3}T_{\rm kd}^{3} t_{\rm eq} G^{7/2}}},\\
    M_{2}& = \frac{1}{8}\sqrt{\frac{45}{\pi^{3}g_{\rm eff}(T_{\rm eq})}} \sqrt{\frac{\eta_{\rm ta}}{m_{\chi}^{3}T_{\rm kd} G^{3}}}.
    \eeq
Thus one can more intuitively understand these three cases as mass regimes. 
Namely, one has a light PBH regime for $M_\bullet<M_1$, a heavy regime for $M_\bullet>M_2$, and an intermediate class for $M_1<M_\bullet<M_2$.  Switch overs in the analytic behaviour of the dark matter halo (governed by $M_1$ and $M_2$) depend also on the properties of the dark matter, as might be anticipated, specifically, the  dark matter  mass $m_{\chi}$, and the point of kinetic decoupling $x_{\rm kd}$.  Below we summarise the analytic PBH halo profiles derived in \cite{Boudaud:2021irr} in each of the three mass regimes.

\subsection{Lighter black holes}\label{2.1}

In the light PBH case with (evaluating the form of eq.~(\ref{M1M2}))
\beq
M_{\bullet} \leq M_{1}\sim3\times 10^{-10}M_\odot\left(\frac{100~{\rm GeV}}{m_\chi}\right)^{3}\left(\frac{x_{\rm kd}}{10^4}\right)^{3/2}~,
\eeq
 the dark matter density profile surrounding the PBH is found to have two scaling regimes, the profile in the innermost region scales as $\rho_{3/4}\propto r^{-3/4}$  and then transitions to a steeper profile with $\rho_{3/2}\propto r^{-3/2}$.
Evaluating eq.~(\ref{eq:DensityMiniSpike}) in the light PBH limit the corresponding dark matter density profile $\rho(r)$ is found to be well described by the following piecewise function
\beq\label{light}
\rho_{\rm light}(\tilde{r})= \left\lbrace
\begin{array}{ll}
~\rho_{3/4}(\tilde{r})&\qquad \tilde{r}< \tilde{r}_A\\[5pt]
~\rho_{3/2}(\tilde{r})&\qquad  \tilde{r}_A<\tilde{r}<\tilde{r}_{T}\\[5pt]
~0&\qquad \tilde{r}> \tilde{r}_T 
\end{array}\right.
\eeq
where $r_A$ corresponds to the characteristic radius of the transition between profiles, and $r_T$ is the terminal extent of the halo. Neglecting stripping of the outer profile by other astrophysical bodies the profile naturally terminates at $r_T = r_{\rm eq}$. However, astrophysical stripping of the outer profile will generically cause the profile to terminate at much small radii, i.e.~$r_T\ll r_{\rm eq}$, we will return to discuss $r_T$ in Section \ref{3.4}. 

We next give the forms of the component-wise dark matter density profiles. In the innermost region the dark matter density has the following radial profile \cite{Boudaud:2021irr}
\beq\label{eq:density34th}
\rho_{3/4}(\tilde{r}) & =\sqrt{\frac{2^{5/2}}{\pi^{3}}}\Gamma(7/4)\mathcal{I}_{3/4}\frac{\rho_{i}^{\rm kd}}{\sigma_{\rm kd}^{3/2}}\tilde{r}^{-3/4},
\eeq
where $\Gamma(x)$ is the Gamma function, and $\mathcal{I}_{3/4}$ is an $\mathcal{O}(1)$ numerical factor coming from evaluating the integral
\beq
\mathcal{I}_{3/4}=\frac{2}{3}\int_{-\infty}^{1}{\rm d}\mathcal{Y}_{m}\frac{\mathcal{F}(\mathcal{Y}_{m})}{\left(1-\mathcal{Y}_{m}\right)^{5/4}}\simeq 4.2.
\eeq
At larger radii the halo profile transition to the following profile  
\beq\label{eq:density32nd}
\rho_{3/2}(\tilde{r}) & = \sqrt{\frac{2}{\pi^{3}}}\frac{\rho_{i}^{\rm kd}}{\sigma_{\rm kd}^{3}}\left(\mathcal{I}_{3/2}  - \frac{3\pi}{8}\frac{\tilde{r}}{\tilde{r}_{\rm eq}}\right)\tilde{r}^{-3/2},
\eeq
where $\mathcal{I}_{3/2}$ is the value of the integration 
\beq\label{a1}
\mathcal{I}_{3/2}=\int_{0}^{\mathcal{R}_{\rm eq}} {\rm d}\mathcal{R}~\mathcal{R}^{-3/2}\int_{{\rm sup}\lbrace 0, (1-\mathcal{R})\rbrace}^{1}{\rm d}u~(1-u)^{3/2}\mathcal{F}(\mathcal{Y}_{m}),
\eeq
evaluated in the limit $\tilde{r}_{\rm eq}\gg \tilde{r}$ one finds $\mathcal{I}_{3/2}\approx1.047$.
The point at which the density profile transitions between the 3/4 and 3/2 profile, which we label as $\tilde{r}_{A}$, can be found by matching eq.~(\ref{eq:density32nd}) and eq.~(\ref{a1}) to obtain
\beq
\tilde{r}_{A} = \frac{\sigma_{\rm kd}^{-1/2}}{2\left(\Gamma(7/4)\right)^{4/3}}\left(\frac{\mathcal{I}_{3/2} }{\mathcal{I}_{3/4} }\right)^{4/3}\simeq 0.08 x_{\rm kd}
\eeq
where we have used that $\sigma_{\rm kd} = (T_{\rm kd}/m_{\chi})^2$.
Note that since these profiles (and the others which follow) depend linearly on $\rho_{i}^{\rm kd}$ as the initial abundance of particle dark matter is varied (or, analogously, the fractional  abundance $f_{\rm PBH}$ which we define shortly), this scales the overall central density of the PBH halos.

\subsection{Intermediate mass black holes}\label{2.2}
In the intermediate range PBH with $M_{1}\leq M_{\bullet}\leq M_{2}$, the PBH halo profiles can be well described by a similar piecewise function
\beq
\rho_{\rm int}(\tilde{r})= \left\lbrace
\begin{array}{ll}
~\rho_{3/4}(\tilde{r})&\qquad \tilde{r}< \tilde{r}_A\\[5pt]
~\rho_{3/2}(\tilde{r})&\qquad  \tilde{r}_A<\tilde{r}<r_{B}\\[5pt]
~\rho_{9/4}(\tilde{r})&\qquad  \tilde{r}_B<\tilde{r}<r_{T}\\[5pt]
~0&\qquad \tilde{r}> \tilde{r}_T
\end{array}\right.~.
\eeq
Observe that this profile exhibits three different regions, and thus two transition lengths $\tilde r_A$ and $\tilde r_B$. The inner region with $\tilde{r}<\tilde{r}_B$ has the same profile as the lighter mass PBH of Section \ref{2.1} and $\rho_{3/4}$, $\rho_{3/2}$, and $\tilde{r}_{A}$ are given by the same algebraic expressions. However, at larger radii the density profile transforms into an $\rho_{9/4}\propto\tilde{r}^{-9/4}$ profile given by
\beq\label{eq:density94th}
\rho_{9/4}\left(\tilde{r}\right) = \frac{\sqrt{128\pi}}{\Gamma^{2}\left(1/4\right)}\rho_{i}^{\rm kd}\tilde{r}_{\rm kd}^{9/4}\tilde{r}^{-9/4}\left[1-\frac{\Gamma^{2}(1/4)}{3\sqrt{2\pi^{3}}}\frac{\tilde{r}^{3/4}}{\tilde{r}_{\rm eq}^{3/4}}\right].
\eeq
As in the previous section, the transition point $\tilde{r}_{B}$ between the $\rho_{3/2}$ profile and the $\rho_{9/4}$ profile can be determined by matching the two profiles.

\begin{figure}[t!]
\centerline{
 \includegraphics[scale=0.4]{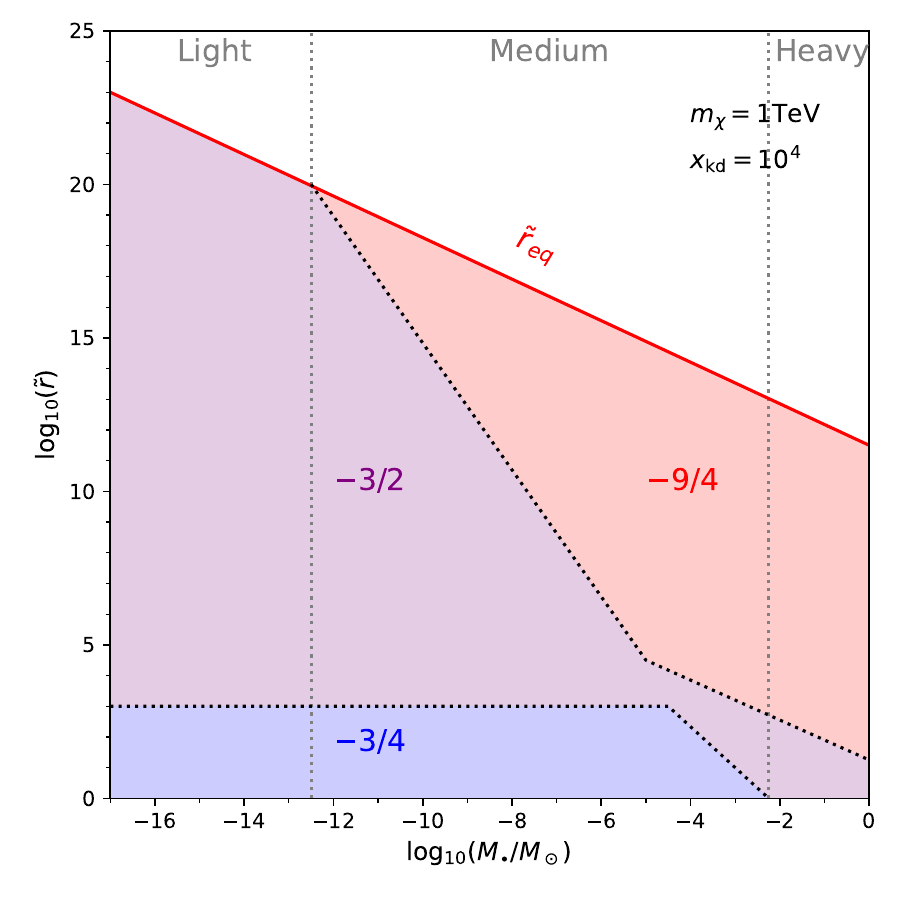}
 \includegraphics[scale=0.4]{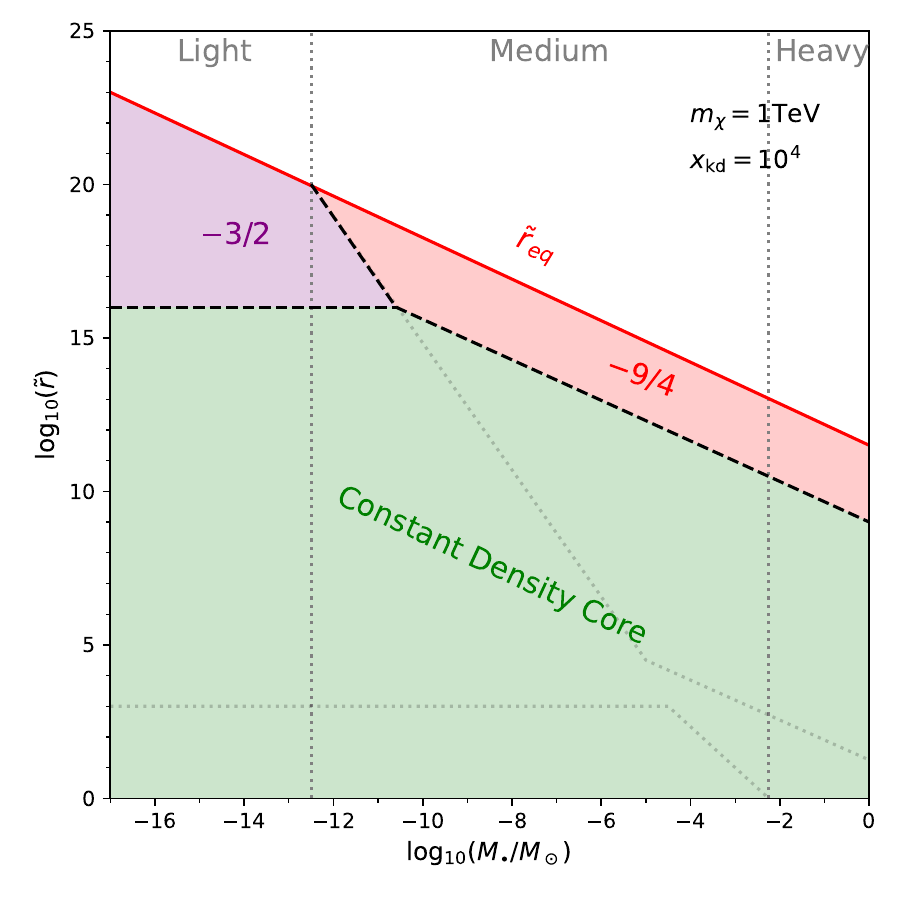}
 \includegraphics[scale=0.4]{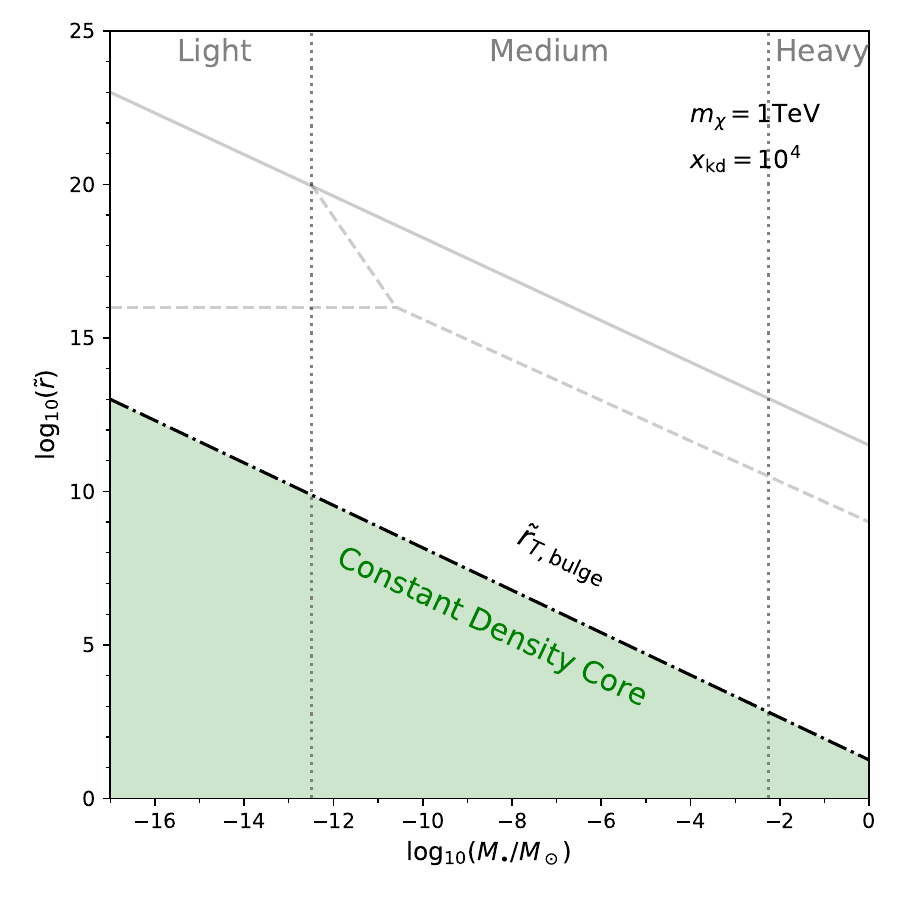}}
\vspace{-3mm}
\caption{
These plots indicate the relevant length scales which dictate the scaling laws of the PBH halo profiles for different PBH masses $M_\bullet$ and under different assumptions  (adopting the style of \cite{Boudaud:2021irr} (Fig.~7)).  Left: We show how the PBH halo profiles vary between the various regimes neglecting particle dark matter annihilations and assuming a halo terminal radius to be $r_T=r_{\rm eq}$, following \cite{Boudaud:2021irr}. Centre: We highlight the impact of including particle dark matter annihilations for $s$-wave annihilations,  as studied in Section \ref{4.2}. Comparing to the left panel we observe that much of the parameter space with non-trivial scaling laws is replaced by a constant central density core. In the case of $p$-wave annihilations, the boundary is shifted by about an order of magnitude and the inner profile is rising instead of constant, however, the qualitative properties of this plot mostly remain unchanged. Right: The PBH profile will be generically stripped through close encounters with astrophysical bodies, the figure shows how the scaling regions are altered assuming that the PBH resides in the Galactic Bulge, see Section \ref{3.4}, note that in this case only the constant density core remains. 
\label{Fig:length}}
\vspace{-3mm}
\end{figure}

\subsection{Heavy black holes}\label{2.3}
For heavy black holes with 
\beq
M_{\bullet}>M_{2}\sim3\times10^{-3}M_\odot\left(\frac{100~{\rm GeV}}{m_\chi}\right)^{2}\left(\frac{x_{\rm kd}}{10^4}\right)^{1/2},
\eeq
 the density profile is well approximated by the following piecewise function
\beq\label{heavy}
\rho_{\rm heavy}(\tilde{r})= \left\lbrace
\begin{array}{ll}
~\rho'_{3/2}(\tilde{r})&\qquad \tilde{r}< \tilde{r}_C\\[9pt]
~\rho_{9/4}(\tilde{r})&\qquad  \tilde{r}_C<\tilde{r}<r_{T}\\[9pt]
~0&\qquad \tilde{r}> \tilde{r}_T
\end{array}\right.~.
\eeq
While the $\rho_{9/4}$ is identical to eq.~(\ref{eq:density94th}), the inner $r^{-3/2}$ profile differs from the case of lighter PBH and follows instead the  form below
\beq\label{eq:density32ndA}
\rho'_{3/2}\left(\tilde{r}\right) = \sqrt{\frac{2}{\pi^{3}}}\rho_{i}^{\rm kd}\frac{\tilde{r}^{3/2}_{\rm kd}}{\tilde{r}^{3/2}}\left(\frac{2\sqrt{2\pi}}{3}\left[2+\left(1+\frac{2\tilde{r}}{\tilde{r}_{\rm kd}}\right)\sqrt{1-\frac{\tilde{r}}{\tilde{r}_{\rm kd}}}\right]\right).
\eeq
Similar to Section \ref{2.1}, the transition from $\rho'_{3/2}$ to $\rho_{9/4}$ is identified by matching the outer and inner profiles and we label this scale $\tilde{r}_{C}^{\prime}$.

We compare our profiles following Boudaud {\em et al.} to other treatments in the literature, specifically Adamek {\em et al.}~\cite{Adamek:2019gns} and Carr {\em et al.}~\cite{Carr:2020mqm} with identical parameter choices. We find that these analytic results are broadly consistent with existing results in the literature. Finally, we note that many studies which aim to constrain dark matter annihilations around PBH simply approximate the profile as $r^{-9/4}$, and thus utilising this careful analytic treatment of Boudaud {\em et al.}~offers an improvement on many earlier approaches. However, as we will see in later sections, particle dark matter interactions and stripping of the halo due to close encounters with astrophysical bodies can significantly alter the halo. 

In Figure~\ref{Fig:length} we summarise the relevant length scales which dictate the scaling laws of the PBH halo profiles. In the left panel we show how the PBH halo profiles vary between the various regimes as described in this section. The centre panel illustrates how the introduction of particle dark matter annihilations alters the picture, as we explain in Section \ref{4.2}. Finally, the right panel shows the potential impact of halo stripping due to encounters with astrophysical bodies, assuming that the PBH resides in the Galactic Bulge, we discuss this in further detail in Section \ref{3.4}. We highlight that for extragalactic $\gamma$-ray constraints we do not anticipate the stripping to be significant and thus the central panel better describes the PBH halo profile for these observations. In contrast, for galactic $\gamma$-ray observations these will be sourced by PBH in the Galactic Bulge at late time and thus the appropriate profiles are better described by the right panel.

\section{The Impact of Dark Matter Interactions on the PBH Halo Profile}
\label{S4}

In the previous section we have outlined the form of the dark matter density profiles around PBH, however it should be stressed that the above dark matter density distributions only hold if the dark matter  has no non-gravitational interactions, or if those additional interactions can be neglected. In particular, both dark matter annihilations and scattering can  potentially alter the halo shape. Notably, if the relic density of particle dark matter is set via freeze-out, as assumed here, then the annihilation rate of dark matter in the centre of the halo is sufficient to considerably alter the halo profiles identified in Section \ref{S2}, as we discuss next.

\subsection{Thermal Dark Matter}

The guiding premise of thermal dark matter is that dark matter was once in thermal equilibrium with the Standard Model states and subsequently decoupled via ``freeze-out'' leading to the observed relic density \cite{Scherrer:1985zt}. 
Since both particle dark matter and PBH contribute to the observed dark matter today we encapsulate their contributions in the form 
\beq
\Omega_{\rm DM}=\Omega_{\rm PBH}+\Omega_{\rm pDM}~.
\eeq
This implies that one can make the replacement  $\Omega_{\rm pDM} = \Omega_{\rm DM}(1-f_{\rm PBH})$ in eq.~(\ref{eq:PreCollapse-density}) and we see that the density of particle dark matter is reduced for larger $f_{\rm PBH}$, as expected. Moreover, we can now formally define PBH fractional abundance as follows
\beq\label{fmax}
f_{\rm PBH}= \frac{\Omega_{\rm PBH}}{\Omega_{\rm DM}} = \frac{\Omega_{\rm PBH}}{\Omega_{\rm PBH}+\Omega_{\rm pDM}}~.
\eeq
In calculating the late time abundance of particle dark matter in this framework it is typical to consider the thermally averaged annihilation cross-section 
\beq
\langle\sigma v\rangle = \sqrt{\frac{x^{3}}{4\pi}}\int_{0}^{\infty}dv~v^{2}e^{-xv^{2}/4}\sigma v,
\eeq
and take a standard expansion of the thermally averaged cross-section
\beq\label{eq:AverageCrossSection}
\langle \sigma v\rangle &= \langle \sigma_s +\sigma_p v^2+\sigma_d v^4+\cdots \rangle~.
\eeq
One can rewrite this expansion in terms of the inverse temperature  \cite{Gondolo:1990dk}
\beq
\langle \sigma v\rangle &=  \sigma_s +\sigma_p  \frac{3}{2x} +\sigma_d  \frac{15}{8x^2}+\cdots 
\eeq
We identify the coefficients to each term in the velocity expansion as the $s$-wave ($\sigma_s$), $p$-wave ($\sigma_p$), and $d$-wave ($\sigma_d$) pieces  to the cross section, being the sequential leading contributions. 
The $s$-wave piece is velocity independent, in contrast to other contributions. Since at freeze-out $x_F\sim30$, each subsequent term exhibits a substantial suppression relative to previous terms in the expansion. Thus if $\sigma_s\neq0$ this contribution will dominate the annihilation cross section, we refer to this as $s$-wave annihilating dark matter. Similarly, if  $\sigma_s=0$ but  $\sigma_p\neq0$ then it is these $p$-wave contributions that dominate the annihilation process, and so on. It is useful to keep in mind that often, even though at tree level $\sigma_s = 0$, there will typically be loop induced contributions to $\sigma_s$ that while subleading at freeze-out, may become relevant for low velocities within the PBH halo and we will discuss the impacts of this in Section \ref{S6}. 

Notably, there is a host of well motivated dark matter models in the literature that lead to $s$-wave and $p$-wave annihilating dark matter scenarios (see e.g.~\cite{Shelton:2015aqa} for $p$-wave models). Given a specific Lagrangian which describes the dark matter interactions one can calculate  the thermally averaged annihilation cross-section, and parameterise it in the form of eq.~(\ref{eq:AverageCrossSection}) and thus identify any velocity suppression in the leading contribution to the annihilation cross section. In Appendix \ref{A3} we sketch a simple model of $d$-wave annihilating dark matter (following \cite{Giacchino:2013bta,Boddy:2019qak}) which reproduces the observed relic density via thermal freeze-out whilst avoiding direct detection constraints.

%
\begin{figure}
\centerline{
 \includegraphics[height=55mm]{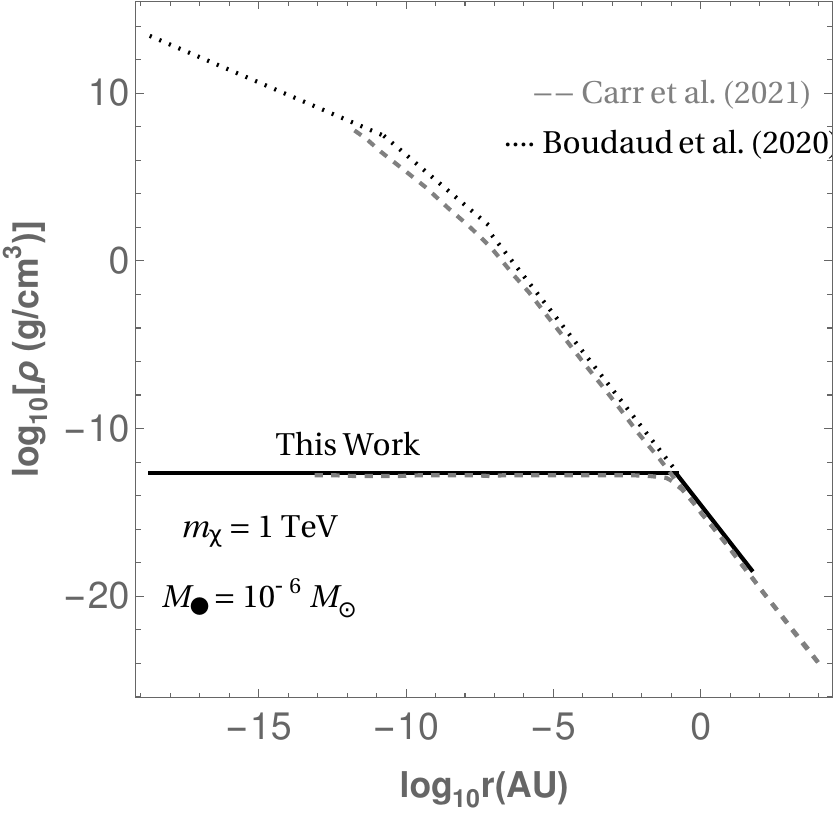}
\hspace{1cm}
 \includegraphics[height=55mm]{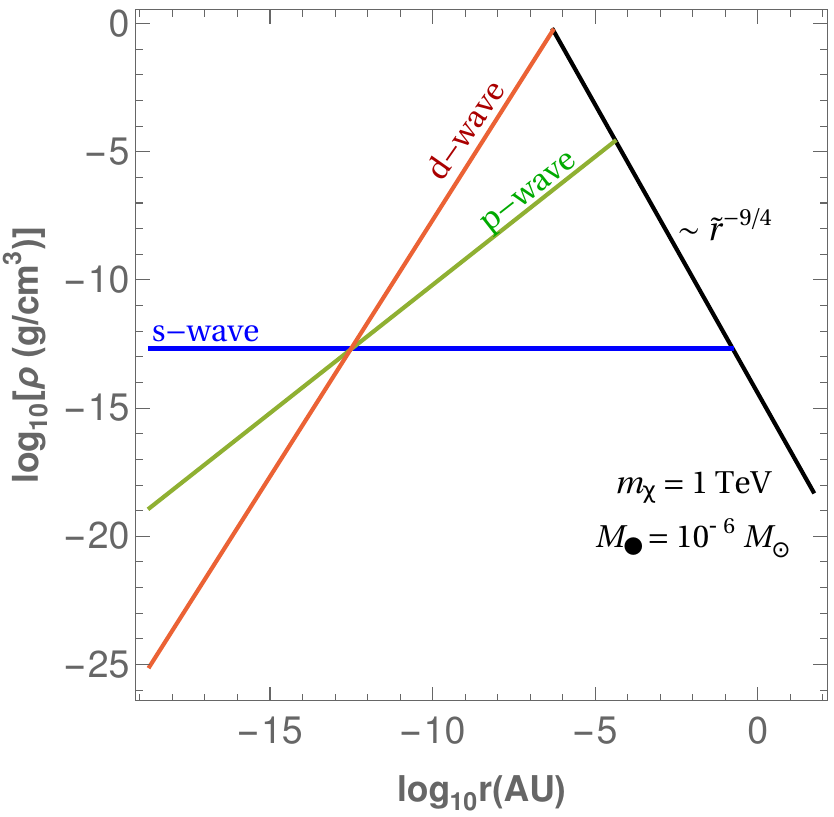}}
 \vspace{-3mm} 
\caption{Left.~The dark matter density profile around the PBH of Boudaud {\em et al.}~\cite{Boudaud:2021irr} (dashed), the impact of including annihilation labelled ``This work'' (solid), and also compare to the alternatively derived profile of Carr {\em et al.} \cite{Carr:2020mqm} which includes dark matter annihilations (grey dashed).  In all cases dark matter annihilation is assumed to be $s$-wave, other parameter values are stated. Observe that with dark matter annihilations the density profile features an inner plateau and then falls with a power-law profile at increasing radial distance. Right.~A comparison of three density profiles with the same parameters, but assuming only $s$-wave, $p$-wave and $d$-wave annihilations. The velocity dependence of the annihilation cross section translates into the radial dependence of the central density profile.
\vspace{-3mm}%
\label{fig:2}}
\end{figure}
%

\subsection{Dark Matter Annihilations}\label{4.2}
Due to the high density environment inside the PBH halo the dark matter particles will undergo annihilation resulting in a flattening  of the inner density spike if the annihilation rate is sufficient. Thus the density near to the center of the halo will be determined by the annihilation rate, which scales as $\Gamma_{\rm ann}\sim n_{\chi}(z)\langle\sigma v\rangle$, where $n_{\chi}(z)$ is the dark matter number density at a given redshift $z$, and $\langle\sigma v\rangle$ is the thermally averaged annihilation cross-section. It follows that the maximum density of dark matter after annihilations will be \cite{Eroshenko:2016yve}
\beq\label{eq:rhomax}
\rho_{\rm max} \approx m_{\chi}n_{\chi}\sim\frac{m_{\chi}\Gamma_{\rm ann}}{\langle\sigma v\rangle}
\sim \frac{m_{\chi}}{\langle\sigma v\rangle t_{\rm halo}},
\eeq
where we take $\Gamma_{\rm ann}\sim 1/t_{\rm halo}$, with $t_{\rm halo}\sim 10^{10}$ years being the age of the halo. However, note for certain applications, like the calculation of the diffuse gamma-ray background, the relevant age of the halo can be significantly lower as we discuss in Section \ref{S5}.

  Accordingly, the halo profiles for dark matter with velocity independent annihilation cross sections ($s$-wave) exhibit central regions for which the density profile is flat with $\rho(r)=\rho_{\rm max}$ for $r<r_{\rm core}$. More generally, for velocity dependent annihilation cross sections $\rho_{\rm max}$ is not constant, but inherits the radial dependance coming from the velocity dependence in the cross section with $v(r)\simeq\sqrt{GM_\bullet/r}$. Beyond $r_{\rm core}$ the density $\rho(r)$ smoothly transitions into the profiles described in Section \ref{S2}. Thus one can determine $r_{\rm core}$ by the matching condition 
  $\rho_{\rm max}(r_{\rm core})=\rho_{i}(r_{\rm core})$  for $i=$light, int, or heavy as given in Section \ref{S2}. 
Figure \ref{fig:2} (left) provides a comparison between the density profile with and without dark matter $s$-wave annihilations, also comparing with the profile of Carr {\em et al.}~\cite{Carr:2020mqm}. Figure \ref{fig:2} (right) gives an example of how  $\rho_{\rm max}(r)$ varies in the cases that the annihilation channel is $s$-wave, $p$-wave or $d$-wave; in particular, observe that the radial dependence in the latter two cases.

Returning to inspect Figure \ref{Fig:length}, the impact of including particle dark matter annihilations is shown in the central panel for $s$-wave annihilations. We highlight that compared to the left panel which neglected dark matter interactions, much of the parameter space in which the halo exhibited an interesting scaling law has now been replaced with a constant central density core. In the next  subsection we proceed to discuss the prospect of halo stripping, corresponding to the right panel of Figure \ref{Fig:length}.

\subsection{Stripping Radius}\label{3.4}

We now discuss the terminal radius of the PBH dark matter halo $r_T$. Much like planets in the outer-reaches of a star system can be stripped away from their parent star if another star passes too close to the system, close encounters of a PBH with astrophysical objects of similar or greater mass can strip the exterior of the PBH's dark matter halo,  such that $r_T\ll r_{\rm eq}$. Close encounters with stars are the obvious candidate for stripping events, however close encounters with the galactic centre, as well as other PBH, can also be relevant.
This discussion is included to convey that the late-time halos around PBH can be quite different from those at earlier times. Notably, we do not expect stripping to impact constraints  from extragalactic gamma-rays (which will be our focus in Section \ref{S5}) since these are dominantly sourced at higher redshift, prior to stripping events. 

Suppose that a PBH undergoes a close encounter with body $B$ with density $\rho_B$.
One can estimate the radius $r_T$ out to which the dark matter halo survives stripping due to a close encounter with $B$ via the Hill radius, this leads to a terminal radius of order (see e.g.~\cite{Scholtz:2019csj})
\beq
r_T=\left(\frac{\rho_{\rm halo}}{2\rho_B}\right)^{1/3}\sim d\left(\frac{M_{\rm halo}}{2M_B}\right)^{1/3},
\eeq
where $d$ is the distance between the PBH and $B$ at the point of closest approach, and $\rho_{\rm halo}$ and $M_{\rm halo}$ are the total density and mass of the PBH halo prior to the close encounter.

 To arrive at an order of magnitude estimate, we suppose that the most significant  tidal stripping events are due to close encounters with stars, thus we take $M_B\equiv M_\odot$. Given that a PBH will have traversed the galaxy for $10^{10}$ years and for the typical spacing of stars we take 0.01~pc,\footnote{The typical spacing of stars is  $\sim$1 pc in the solar neighbourhood and $\lesssim0.01$ pc in the buldge, see e.g.~\cite{Balbi}} taking the typical PBH speed to be $\sim200$~km/s (in line with the typical dark matter velocity) then each PBH will have encountered around $N_\star$ stars:
\beq
N_\star \sim \mathcal{O}(10^8)
\eeq
 Note that if the PBH have orbits such that they never pass through the Galactic Bulge then the number of star encounters drops to $\mathcal{O}(10^6)$, however, we consider here the more aggressive scenario since this would have the biggest impact on our calculations.

If make an large simplification and suppose that the stars in the galaxy roughly form a regular square lattice with $l = 10^{-2}$pc spacings, then it follows that the closest encounter of a PBH with a star located at one of these lattice sites is roughly 
\beq
d\sim \frac{l}{\sqrt{N}} = 10^{-6} \rm pc.
\eeq
 It follows that during the history of the PBH one anticipates that  such close encounters with stars should truncate the PBH's associated dark matter halo to around
\beq\label{eq:rT}           
  r_T(\mathrm{bulge}) \sim d\left(\frac{M_{\rm halo}}{2M_\odot}\right)^{\frac{1}{3}}\sim  10^{-6}\mathrm{pc}
              \left(\frac{d}{10^{-6}~\mathrm{pc}}\right)  \left(\frac{M_{\mathrm{halo}}}{1 M_\odot}\right)^{\frac{1}{3}}.
\eeq
%
Outside of the bulge, where encounters are less common, this can be a much larger scale:
$l = 1\rm pc$, so $N \sim \mathcal{O}(10^6)$, hence $d = 10^{-3} \mathrm{pc}$ and so:
\beq\label{eq:rTdisk}           
  r_T(\mathrm{disk}) \sim  10^{-3}\mathrm{pc}
              \left(\frac{d}{10^{-3}~\mathrm{pc}}\right)  \left(\frac{M_{\mathrm{halo}}}{1 M_\odot}\right)^{\frac{1}{3}}.
\eeq
%
Furthermore, PBHs that are in the dark matter  halo, have significantly suppressed encounter rate (only when they cross the disk, twice per their orbit of the galaxy) and so their terminal radius can be even larger.
 
For the total mass of the PBH halo prior to the close encounter $M_{\rm halo}$ we have normalised the expression to $M_\odot$. 
Moreover, in Appendix \ref{A1} we calculate the typical total mass of the PBH halo due to accretion and also following annihilations. In particular, we find that  $M_{\rm halo}\sim M_\bullet$ is a characteristic total mass for the PBH dark matter halo and thus $  r_T\sim  10^{-6}\mathrm{pc}$ is appropriate for  $M_\bullet\sim 1M_\odot$.
Furthermore, examining Figure \ref{Fig:length} again, we note that the right panel illustrates the case that the terminal radius of the halo is identified as $  r_T(\mathrm{bulge})$, whereas the left/centre panels assume   $r_T=r_{\rm eq}$. Observe that the change in the stripping radius significantly impacts the expectation for the late time PBH halo profile. In Appendix \ref{Anew} we also discuss the possibility of PBH stripping due to interactions with other PBH.

While this estimate of the terminal radius is a rather quick analysis, the exact details are largely unimportant for our purposes since the vast majority of dark matter annihilations will occur in the high-density central region. Moreover, the precise radius at which the halo terminates will have little impact on the leading constraints coming from extragalactic $\gamma$-ray observations, as we discuss in the next section.  

\section{Constraints on PBH Dark Matter Halos}
\label{S5}

The dense dark matter halo around the PBH, as detailed in the previous sections, will invariably lead to the production of high energy photons and other observable particles if the dark matter has any appreciable couplings with Standard Model states. This observation has been used in earlier studies to derive stringent constraints on $s$-wave annihilating WIMP dark matter; here we extend this line of reasoning to place constraints on dark matter in the case that the relic density is due to thermal freeze-out via $p$-wave or $d$-wave annihilation processes. Moreover, in deriving our bounds on dark matter annihilations we employ the more sophisticated halo profile developed in~\cite{Boudaud:2021irr} and outlined in Section \ref{S2}.

\subsection{Annihilations in the PBH Halo}

If the relic abundance of particle dark matter is set via thermal freeze-out then the dark matter will have appreciable couplings to the Standard Model, it follows that dark matter states may continue to annihilate at the present day in regions of enhanced density. In particular, the highly dense halos surrounding PBH are ideal environments for particle dark matter annihilations. Such dark matter annihilations can lead to observable signals, in particular $\gamma$-rays. As a result, we can place limits on the combined system of dark matter plus PBH from the null observations in searches for extragalactic  $ \gamma$-ray excesses.

We shall work in a model independent fashion,  thus we identify the value of $\sigma_i$ (with $i=s,p,d$) required to reproduce the observed dark matter relic density $\Omega_{\rm DM}\approx0.26$, such that $\sigma_i\neq0$ is the leading non-zero term in the expansion of eq.~(\ref{eq:AverageCrossSection}). For the classic $s$-wave annihilating ``WIMP''  this is achieved for $\sigma_s\sim3\times10^{-26}{\rm cm}^3/{\rm s}$ for Weak scale dark matter masses (this value is only logarithmically sensitive to changes in the mass) provided that particle dark matter is mostly responsible for inferred value $\Omega_{\rm DM}\approx0.26$. The cross section for $p$-wave and $d$-wave dark matter to match the observed relic density will be numerically different, since it is suppressed by powers of the velocity, but can be calculated similarly.

In our calculations we fix the value of $\sigma_i$ by requiring that $\Omega_{\rm pDM}=\Omega_{\rm DM}\approx0.26$. This is  technically only correct for $f_{\rm max}\ll1$ since in this mixed PBH-particle dark matter scenario the PBH can account for an appreciable, or even the dominant, contribution to the observed quantity $\Omega_{\rm DM}=\Omega_{\rm pDM}+\Omega_{\rm PBH}$. However, the impact to the value of $\sigma_i$  required to obtain the correct $\Omega_{\rm DM}$ taking into account $\Omega_{\rm PBH}\neq0$ will be negligible away from $f_{\rm PBH}\approx1$. To circumvent this issue we restrict our analysis to $f_{\rm PBH}\leq0.3$. In addition to making the analysis somewhat simpler, this cut on $f_{\rm PBH}$ is very sensible since over the majority of parameter space fractional abundance $f_{\rm PBH}>0.3$ are excluded by observational constraints~\cite{Carr:2020xqk}.

\begin{figure}[t!]
\centerline{ \includegraphics[scale=0.44]{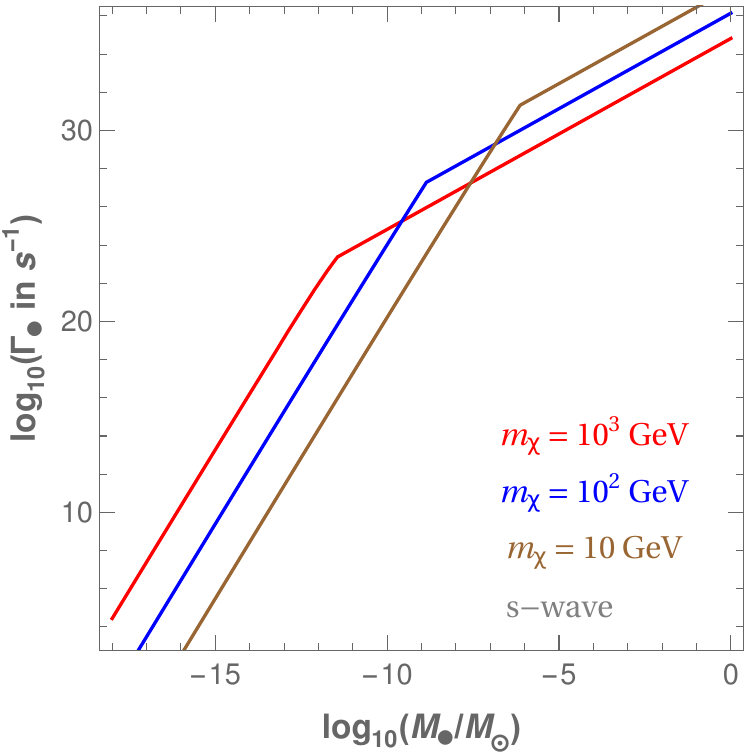}~~
 \includegraphics[scale=0.44]{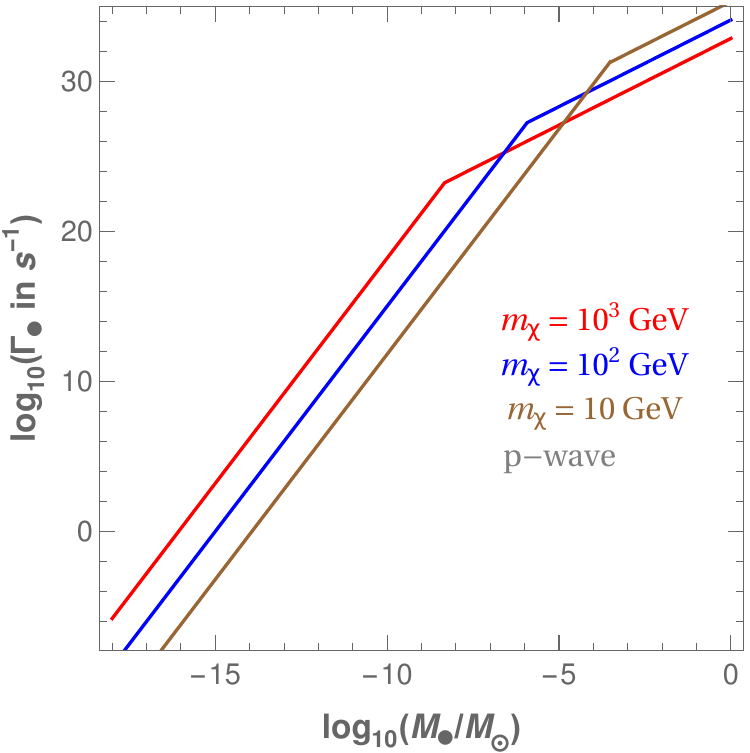}~~
 \includegraphics[scale=0.44]{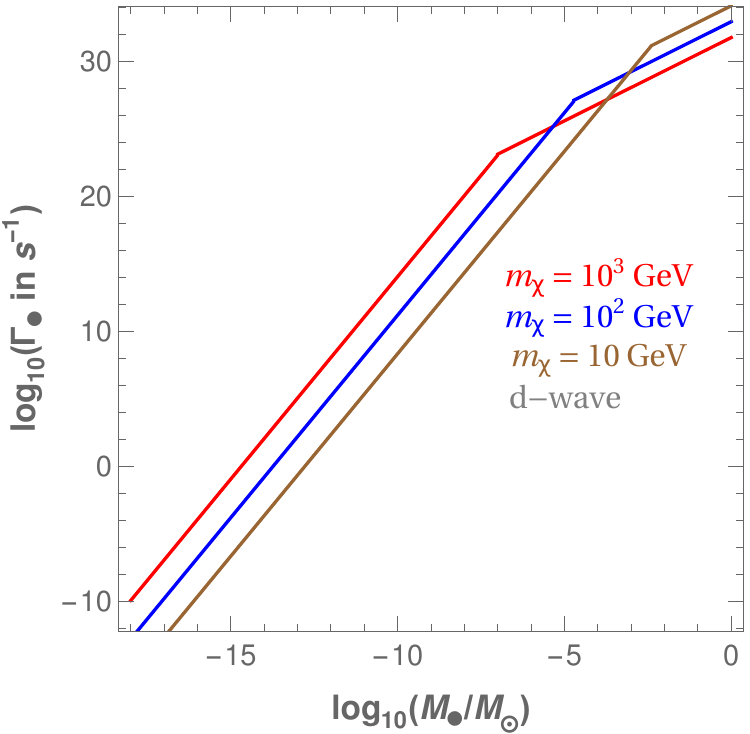}}
\caption{
The annihilation rate, $\Gamma_{\bullet}$, as function of PBH mass $M_{\bullet}$  with a halo extending to $r_T\sim r_{\rm eq}$. We plot $\Gamma_{\bullet}$ assuming that the dominant annihilation channel is either $s$-wave (left), $p$-wave (center), or $d$-wave (right) and for each we show three different dark matter mass choices as indicated. We assume the annihilation rate sets the relic density of particle dark matter $\Omega_{\rm pDM}$ with $\Omega_{\rm pDM}\approx0.26$. As a cross-check, notice that the powerlaw breaks for $m_\chi = 1$~TeV  in the s-channel plot at $M_{\bullet} \sim 10^{-11} M_\odot$, corresponding to the switch over for the outer density profile apparent in the central plot of Figure \ref{Fig:length}.} 
\label{Fig:AnnihilationRate}
\end{figure}

The mixed PBH-particle dark matter scenario necessarily implies a number of free parameters, even in its simplest setting, namely one has the dark matter mass $m_\chi$, the cross section $\langle \sigma v\rangle$, the PBH mass $M_\bullet$ (assuming a uniform mass spectrum\footnote{One could consider PBH which are not produced with a uniform mass scale, this would lead to variations in the exact constraints, but is unlikely to impact the broad conclusions.}) and the fractional abundance of PBH $f_{\rm PBH}$. One  should also specify or identify  the point of kinetic decoupling which can impact the form of the halo, as in eq.~(\ref{M1M2}), here we take $x_{\rm kd}=10^4$. Also, we fix the annihilation cross section $\langle \sigma v\rangle$ by requiring that $\Omega_{\rm pDM}\approx0.26$, as discussed previously. For the three remaining parameters the typical way to present limits in this setting is to constrain $f_{\rm PBH}$ as a function of PBH mass $M_\bullet$ for a given dark matter mass~$m_\chi$.

Given the annihilation cross-section $\langle\sigma v\rangle$ and a dark matter density profile $\rho(r)$, one can calculate an annihilation rate $\Gamma_{\bullet}$ for a given object by integrating over the profile \cite{Cirelli:2010xx}
\beq\label{eq:AnnihilationRate0}
\Gamma_{\bullet} &= 4\pi \int dr~r^{2}\left(\frac{\rho(r)}{m_{\chi}}\right)^{2}\langle\sigma v\rangle ~.
\eeq
In Figure \ref{Fig:AnnihilationRate} we show the variation in the annihilation rate $\Gamma_\bullet$ from the PBH dark matter halo as a function of the PBH mass $M_\bullet$, integrating up to $r=r_{\rm eq}(M_\bullet,x_{\rm kd})$ and we take $x_{\rm kd}=10^4$. The annihilation rate is shows for three different dark matter mass choices and assuming that the dominant annihilation channel is either $s$-wave (left), $p$-wave (center), or $d$-wave (right).
From the annihilation rate $\Gamma_{\bullet}$ one can place limits on $f_{\rm PBH}$ for a given dark matter model by comparing to $\gamma$-ray searches.

\subsection{The Extragalactic $\gamma$-ray Flux due to Annihilations in PBH Halos}

We now examine the limits that come from null searches for excesses of $\gamma$-rays of extragalactic origin, we note that the physics is much richer for the case of velocity dependent dark matter annihilations. 
The rate  of dark matter annihilations $\Gamma_{\bullet}$ is a function of the annihilation cross section $\langle\sigma v\rangle$ (cf.~eq.~(\ref{eq:AnnihilationRate0})), which itself  will depend on the dark matter mass $m_{\chi}$, the dark matter coupling, and, for $p$/$d$-wave annihilation processes, the dark matter velocity. Notably, the velocity of the dark matter is dictated by its orbit around the PBH (and distinct from the velocity used in freeze-out calculations). For calculating the cross section for dark matter annihilations within a halo via the velocity expansion in eq.~(\ref{eq:AverageCrossSection}) we take $v(r)=\sqrt{GM_\bullet/r}$, so that the dark matter velocity varies with its radial distance from the PBH.\footnote{Close to the Schwarzschild radius $r\sim\mathcal{O}(1) r_{\rm sch}$ the dark matter will reach velocities $v\sim c/\mathcal{O}(1)$ however this is a small fraction of the dark matter and thus we can neglect relativistic corrections to the velocity.}  

For extragalactic $\gamma$-rays the halo density profile is redshift $z$ dependent, hence the effective rate of annihilations is $z$-dependent. Thus we define $\hat\Gamma[\Gamma,h(z)]$ which we write as a function of $h=H(z)/H_0$ and $\Gamma_{\bullet}$  the (potentially velocity dependent) annihilation rate of dark matter in the PBH halo, as given in eq.~(\ref{eq:AnnihilationRate0}). The redshift enters in such a manner that $\hat\Gamma_{\bullet}=\Gamma_{\bullet}P[h(z)]$ where $P$ is some polynomial or logarithmic function determined by the size of the PBH and the annihilation channel, and is such that $\hat\Gamma_{\bullet}|_{z=0}=\Gamma_{\bullet}$. In the  analysis of \cite{Carr:2020mqm} the $z$-dependence of the annihilation rate was taken to be 
\beq\label{GB}
\hat\Gamma_{\bullet,~s}^{B}(z) = \Gamma_{\bullet}(h(z))^{2/3},
\eeq 
where for  $\hat\Gamma_{\bullet,~s}^{B}$ we introduce the subscript $s$, since this form is only valid for $s$-wave annihilation, and the superscript $B$ (for Bertschinger \cite{Bertschinger1985}) since this form is correct only for an $r^{-9/4}$ profile. The study of \cite{Gines:2022qzy} used a rescaling to adjust for this variation in redshifting, but did not take into account the full form of $z$-dependance, as we do here.
Thus with velocity dependent cross sections, or away from the heavy PBH limit $M_{\bullet}>M_{2}$ the form of $\hat\Gamma_{\bullet}$ given in eq.~(\ref{GB}) no longer holds.

For realistic halos (away from the simple $r^{-9/4}$ profile) the $h(z)$ dependance is quite complicated, and
the full forms for the $z$-dependence of the annihilation rate are presented in Appendix  \ref{A4} for each of the various cases. Thus our analysis of the extragalactic $\gamma$-ray bounds improves earlier studies in the literature, even in the well studied $s$-wave case.
   In Appendix  \ref{A4} we give the approximate dependence of the annihilation rate on the redshift, however let us highlight some general features here. For lighter ($L$) black holes $\hat\Gamma_{\bullet}^{L}$ (those of Section \ref{2.1}), and for heavier ($H$) black holes $\hat\Gamma_{\bullet}^{H}$ (those of Section \ref{2.3}), the leading dependance scales with redshift $z$ as: 
   \beq
  \hat \Gamma_{\bullet}^{L}\propto\begin{cases}
1+\mathcal{O}({\rm log}[h(z)]) &   s{\rm-wave} \\[5pt]  
  h^{2/5}(z)  &  p{\rm-wave}  \\[5pt]   
  h^{4/7}(z)   &  d{\rm-wave}
   \end{cases}
   ~~,
   \hspace{2cm}
   \hat\Gamma_{\bullet}^{H}\propto\begin{cases}
h^{2/3}(z) &   s{\rm-wave} \\[5pt]  
  h^{10/13}(z)  &  p{\rm-wave}  \\[5pt]   
  h^{14/17}(z)   &  d{\rm-wave}
   \end{cases}.
   \eeq
The leading dependance is slightly more complicated in the case of intermediate mass PBH (c.f.~Section \ref{2.2}) and depends upon whether the core due to annihilations terminates at the 3/2 profile or the 9/4 profile. In the analysis of \cite{Carr:2020mqm} it would appear that the rate $\hat\Gamma_{\bullet,~s}^{B}(z) = \Gamma_{\bullet}(h(z))^{2/3}$  appropriate for heavier PBH is used for  regardless of PBH mass. 

   Annihilations of the dark matter particles in the halos around PBHs produce radiation in $\gamma$-rays which contribute to the extragalactic differential flux \cite{Ullio:2002pj} 
\beq\label{eq:ExGal-differential-flux}
\left.\frac{{\rm d}\Phi_{\gamma}}{{\rm d}E {\rm d}\Omega }\right|_{\rm ExGal} = \int_{0}^{\infty}dz~
\frac{\hat\Gamma_\bullet(z)n_{\rm PBH}}{8\pi H(z)}e^{-\tau (z,E')}\frac{{\rm d}N_{\gamma}}{{\rm d}E'} ~,
\eeq
where $n_{\rm PBH}$ is the number density of PBHs, and $\tau$ is the optical depth at redshift $z$ for photon-matter pair production, photon-photon pair production, and, photon-photon scattering.  To incorporate the optical depth we follow the treatment in \cite{Cirelli:2010xx}.\footnote{While one may be concerned that the optical depth might be altered by baryonic accretion around the PBH, \cite{Lacki:2010zf} highlights that changes to the optical depth should be negligible for sub-steller mass PBH.} The energy spectrum is ${\rm d}N_{\gamma}/{\rm d}E'$ and we use a prime to indicate that this is the energy at which the annihilations occur, which (due to redshifting) need not be the energy at which they are observed. 
%
%

\subsection{Constraints from the Extragalactic $\gamma$-ray Background}

For a given PBH mass, dark matter mass, and annihilation cross section there is a maximum fractional abundance $f_{\rm PBH}=f_{\rm MAX}$ for which the PBH contribution to the total  differential extragalactic flux saturates the current experimental limit. Here we compute the constraints by requiring that the PBH contribution to the total  differential extragalactic flux does note exceed the extragalactic $\gamma$-ray background observed by Fermi-LAT \cite{Fermi-LAT:2015qzw}. 

We note that one of the  recent analyses \cite{Carr:2020mqm} to place constraints on $f_{\rm MAX}$ following a similar method seems to have propagated an error from earlier in the literature.  It was shown that the resulting bounds are overstated by several orders of magnitude (especially for lighter dark matter) \cite{Gines:2022qzy}. Our results are more consistent with \cite{Gines:2022qzy} in cases where they overlap, with some differences arising due to slight differences in our analyses. In Appendix \ref{A7} we give a detailed comparison of our work with \cite{Carr:2020mqm}  and \cite{Gines:2022qzy}.

\begin{figure}[t!]
\centerline{
 \includegraphics[scale=0.58]{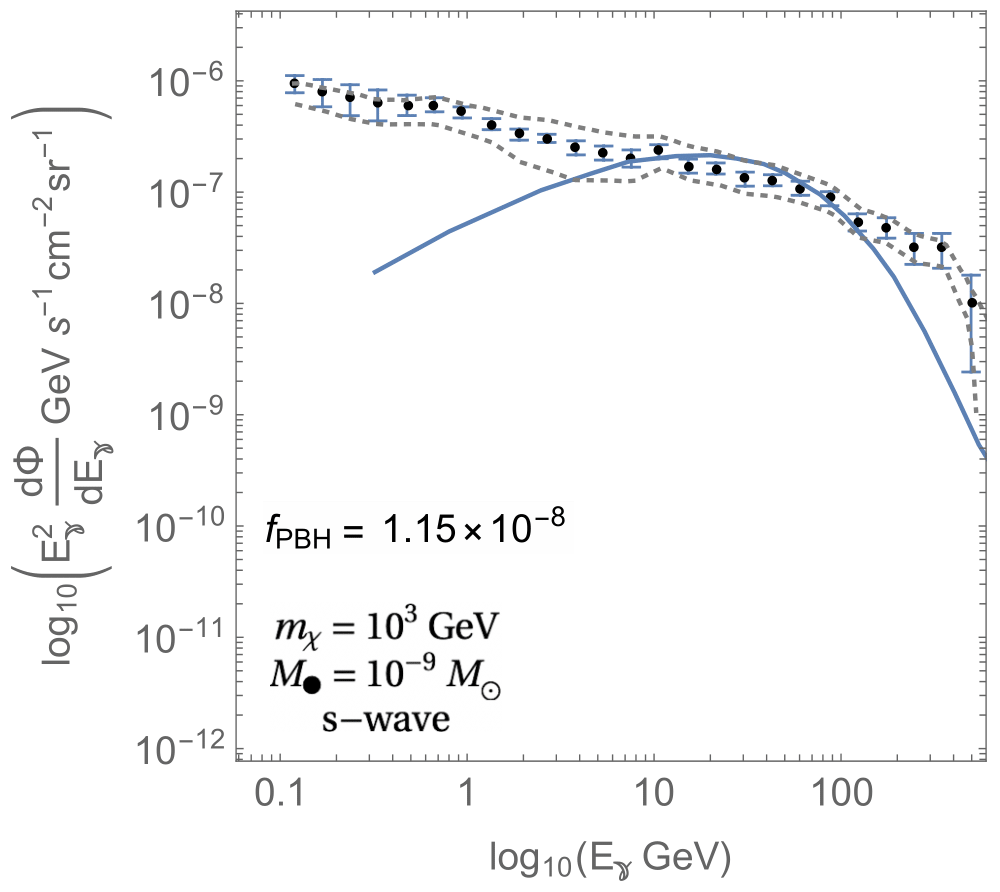}
 \hspace{1cm}
 \includegraphics[scale=0.58]{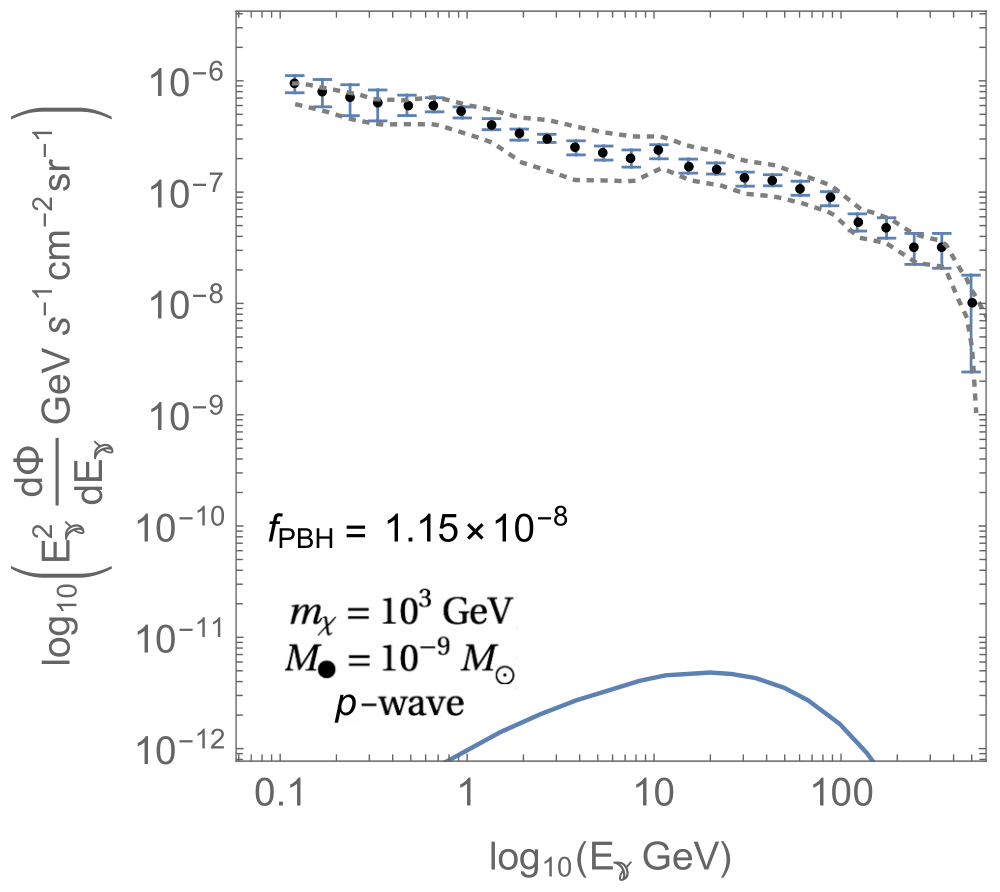}}
\vspace{-3mm}
\caption{Examples of the extragalactic differential flux as a function of energy for dark matter with $m_\chi = 1$~TeV  and PBH mass of  $M_{\bullet} \sim 10^{-9} M_\odot$ for $s$-wave  (left) and  $p$-wave annihilation (right, note different $y$-axis scale), fixing $f_{\rm PBH}=2.8\times10^{-8}$.  We overlay this with the Fermi-LAT data extragalactic $\gamma$-ray background \cite{Fermi-LAT:2015qzw}. By design, for $s$ wave $f_{\rm PBH}=f_{\rm MAX}$ (as defined by our criteria) and one can see that the flux in this case saturates the upper limit of the observed $\gamma$-ray background.
\label{Fig:new}}
\end{figure}

For an analytic intuition for the differential flux, rather than integrating over redshift $z$, we sum over logarithmic spatial shell of redshift $10^{n-1}<z<10^{n}$ (for $n\in\mathbb{N}$): 
\beq\label{eq:ExGal}
\frac{d\Phi_{\gamma}}{dE}\Big|_{\rm ExGal} \simeq \sum_{n=0}^{\infty}10^{n}\times n_{\rm PBH}\left[\frac{dN_{\gamma}}{dE}\frac{e^{-\tau(z,E(1+z))}}{(1+z)H(z)}\hat{\Gamma}_{\bullet}(z) \right]_{z=3\times10^{n}}.
\eeq
The factor $10^{n}$ approximates the interval in $z$ and in each shell one can evaluate the expression in the brackets for the redshift value given by the geometric mean of the boundary values that define the shell $z=\sqrt{10^{n}.10^{n-1}}\approx3\times10^{n-1}$.
For successive spatial shells grow in volume and  the number of PBH in each shell increases exponentially with $n$ and the energy of photons emitted with energy $E$ in shells with $n\gg1$ will be redshifted such that the observed energy is $E = E^{\prime}(1+z)^{-1}$. For the spectra ${\rm d}N_{\gamma}/{\rm d}E$ we use the benchmark models of photon spectra developed in \cite{Cembranos:2010dm} assuming $b\bar{b}$ pairs account for 100\% of the primary annihilation products in calculating the energy distribution.

To obtain reasonable bounds on the differential flux due to annihilation in the PBH halo we bin the flux coming from eq.~(\ref{eq:ExGal}) (after $z$-shell summation/integration) and identify the bin (with central value $E_{\rm peak}$) for which the differential flux is highest.  Notably, the redshifting effects will typically smear the differential flux compared to the expected flux of nearby (i.e.~galactic) populations and may alter the peak. We then impose the simple (but sufficient for our purposes) constraint that 
\beq\label{limit}
\left[ \frac{d\Phi_{\gamma}}{dE}\Big|_{\rm ExGal} \right]_{E=E_{\rm peak}} ~~\leq~~ \left[ \frac{d\Phi_{\gamma}}{dE}\Big|_{\rm Fermi-LAT} \right]_{E=E_{\rm peak}}
\eeq
where ${\rm d}\Phi_{\gamma}/{\rm d}E|_{\rm Fermi-LAT}$ is the observed Fermi-LAT \cite{Fermi-LAT:2015qzw} extragalactic $\gamma$-ray background. 

For a given $E=E_{\rm peak}$ the RHS is simply a number and specifically we assume the value given by the upper limit of their systematic uncertainties. The LHS corresponds to the calculated  differential flux so depends on a number of factors, in particular the dark matter mass, the PBH mass, the annihilation cross-section/channel (final states and Lorentz structure), and the fractional abundance $f_{\rm PBH}$. As noted, we assume that $b\bar{b}$ pairs account for 100\% of the primary annihilation products, and we will consider $s$, $p$, and $d$ wave in turn. 

\begin{figure}[t!]
\centerline{
 \includegraphics[scale=0.44]{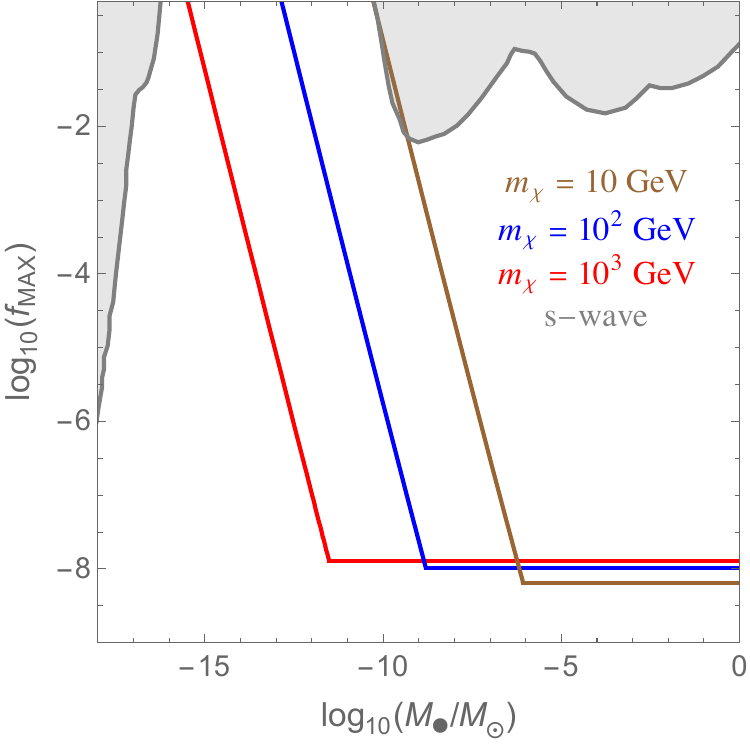}
 \includegraphics[scale=0.44]{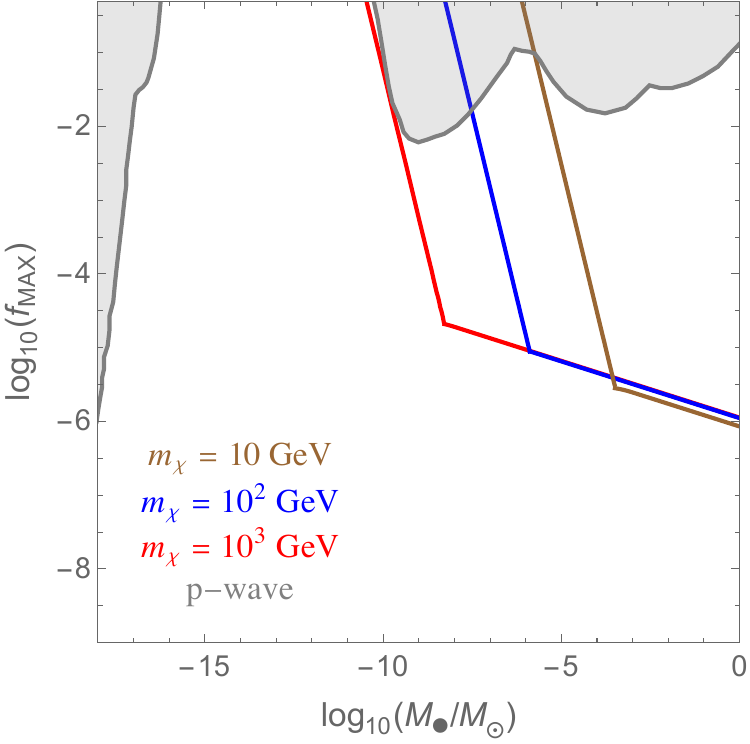}
 \includegraphics[scale=0.44]{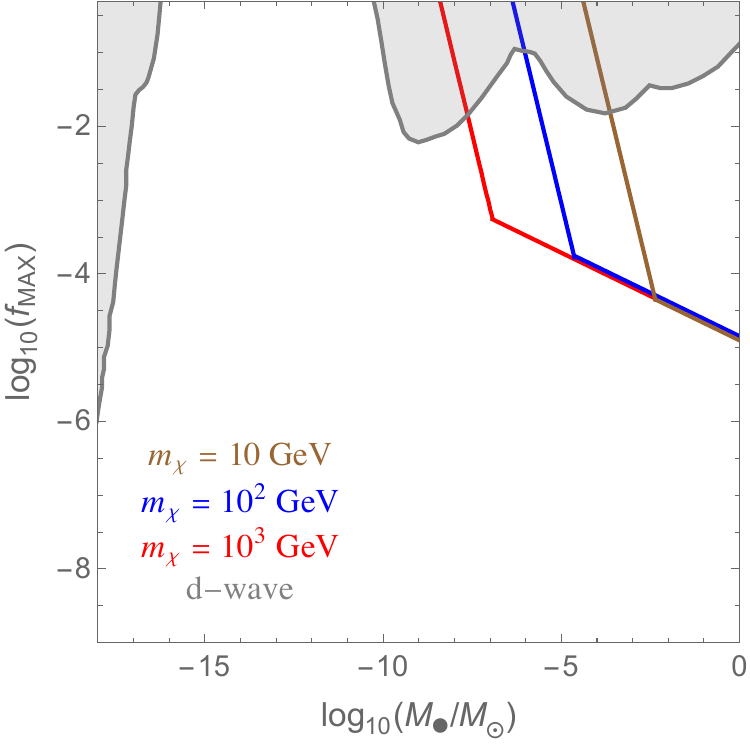}}
\vspace{-3mm}
\caption{The upper bound on the fractional abundance of PBH $f_{\rm PBH}\lesssim f_{\rm MAX}$ from extragalactic $\gamma$-rays as the PBH mass $M_\bullet$ is varied for different dark matter scenarios. We show three choices of the dark matter mass, assuming that the dominant annihilation channel is $s$-wave (left), $p$-wave (centre), and $d$-wave (right). We assume that the same annihilation channel sets the relic density of the particle dark matter. The grey region indicates $M_\bullet-f_{\rm PBH}$ parameter space for which PBH populations are excluded by evaporations, lensing, gravitational waves, or distortions of the CMB (see e.g.~\cite{Carr:2020xqk}).
\label{Fig:sWave-fboundMBH-Galactic}}
\vspace{-3mm}
\end{figure}

Thus for a given dark matter mass and PBH mass one can identify the fractional abundance that saturates the constraint of eq.~(\ref{limit}) which we label $f_{\rm MAX}$. Put another way, for $f_{\rm PBH}>f_{\rm MAX}$  the flux of photons due to annihilations in PBH halos  exceeds the observed extragalactic background.\footnote{One could certainly take a more sophisticated approach by summing over energy bins (see \cite{Cirelli:2012ut} for analyses of this type and discussion), however, the method outlined above is sufficient for our purposes.}
To provide further intuition, in Figure \ref{Fig:new} we present examples of the extragalactic differential flux ${\rm d}\Phi_{\gamma}/{\rm d}E$ for specific values of the dark matter and PBH mass in which we fix $f_{\rm PBH}=f_{\rm MAX}$. We then compare this to the extragalactic $\gamma$-ray background due to Fermi-LAT \cite{Fermi-LAT:2015qzw}, observed that, since we have set $f_{\rm PBH}=f_{\rm MAX}$, the peak of predicted differential flux reaches the upper systematic uncertainty of the extragalactic $\gamma$-ray background. Subsequently, in Figure \ref{Fig:sWave-fboundMBH-Galactic} we show bounds on $f_{\rm MAX}$ as relevant parameters are varied. To determine $E_{\rm peak}$ in Figure \ref{Fig:new} we numerical integrate (rather than summing $z$-shells) from $z=10$ the point of galaxy formation to $z=10^6$ and check that our results are insensitive to order of magnitude variations in the upper limit. Observe that these bounds differ depending on the PBH mass, the dark matter mass, and the annihilation rate (which differs for $s$/$p$/$d$-wave thermal particle dark matter).

\begin{figure}[t!]
\centerline{
 \includegraphics[scale=0.5]{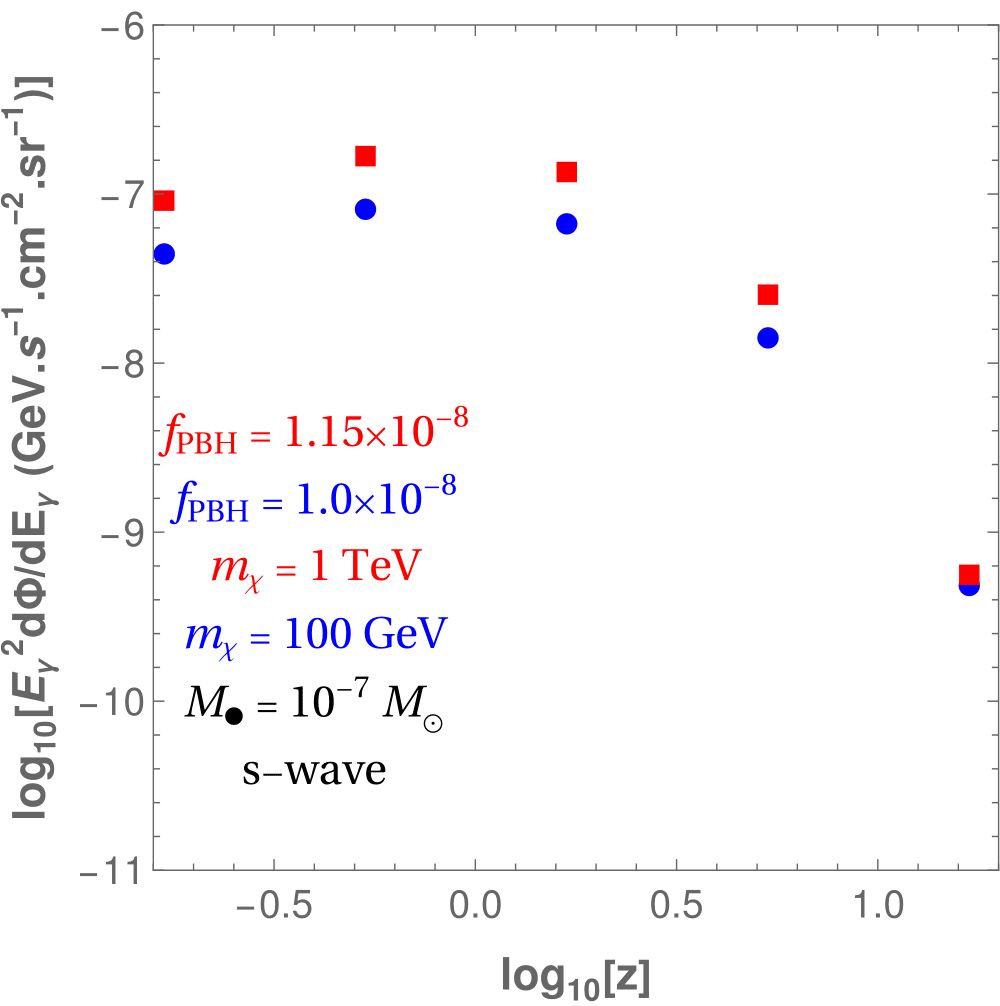} \hspace{15mm}
 \includegraphics[scale=0.5]{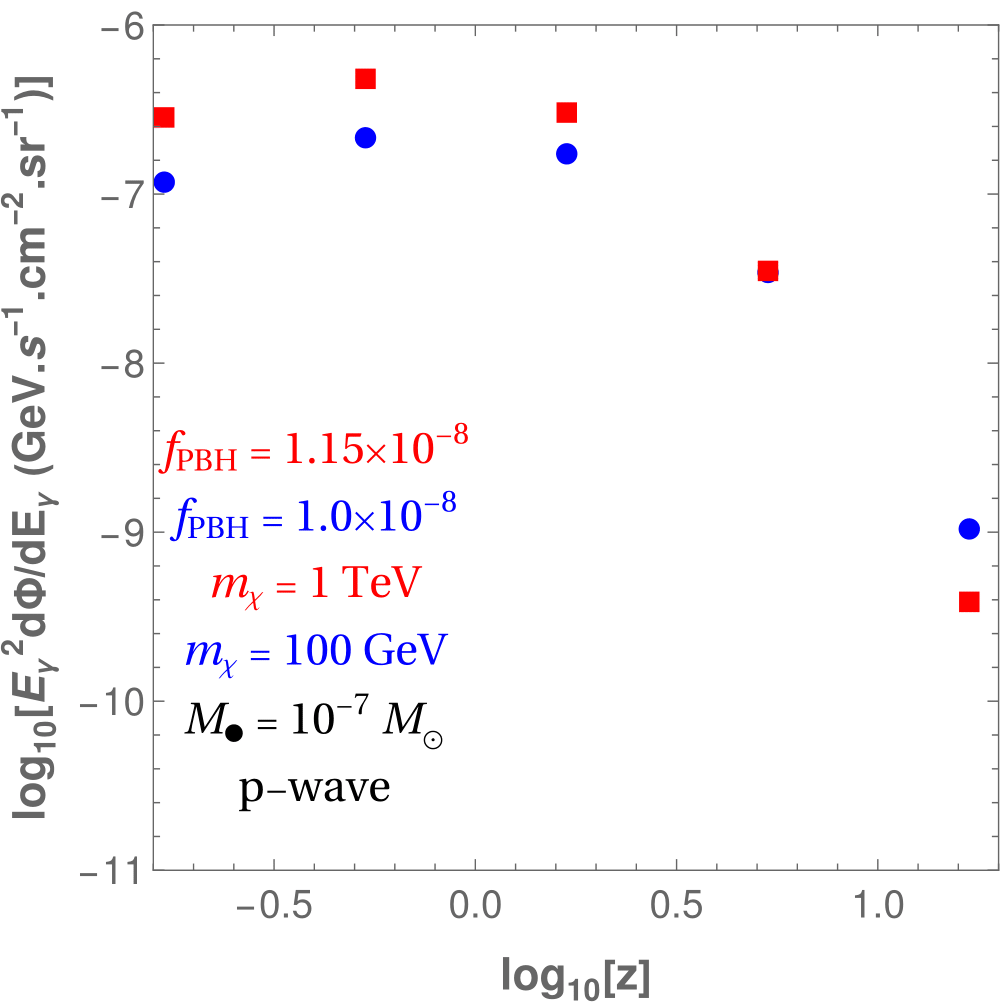}}
\vspace{-2mm}
\caption{Peak value of the extragalactic flux $E^2{\rm d}\Phi/{\rm d}E$ due to sources at different redshift $z$, calculated using successive shells of redshift similar to eq.~(\ref{eq:ExGal}). The  $z\lesssim1$ contribution is subdominant to the sum over higher $z$ contributions. This implies that it is reasonable to neglect the impact of dark matter halo stripping, which will occur for $z\sim1$, in deriving the extragalactic $\gamma$-ray limits.}
\vspace{-2mm}
\label{Fig:newfigure}
\end{figure}

We highlight that the extragalactic bounds are dominated by young PBHs and as a result are not as sensitive to the physics associated with the halo stripping or the dark matter profile changes due to annihilations, as there simply was less time to effect such changes to their profiles. The evolution of the profile due to annihilations is taken into account when integrating over redshift $z$ to compute the extragalactic flux. Stripping likely becomes significant for $z\sim\mathcal{O}(1)$ and for higher-$z$ PBH one can reasonably take the terminal radius of the halo to be $r_T\simeq r_{\rm eq}$,  as we do in calculating the bounds from extragalactic $\gamma$-rays. In Figure \ref{Fig:newfigure} we confirm that the extragalactic bounds are  dominated by the summed contributions from  PBHs at $z\gtrsim1$. In contrast, $\gamma$-rays from galactic sources (as considered in e.g.~\cite{Carr:2020mqm}) are sourced from late-time PBH, and likely dominated by those in the Galactic Bulge,  in which case the relevant PBH halo will have undergone significant stripping one expects $r_T\sim r_{\rm bulge}\ll r_{\rm eq}$. Thus the terminal radius of the PBH relevant to the galactic $\gamma$-rays bounds should leave only a small fraction of the original profile and this will significantly impact the exclusion limits. Accordingly, here we focus on the extragalactic bounds, but moreover it has previously been suggested that bounds derived from galactic populations of PBH tend to be weaker even not accounting for halo stripping \cite{Carr:2020mqm}.

Finally, we note that groups have endeavoured to place other bounds on PBH with dark matter halos, though CMB studies \cite{Gines:2022qzy} or recasting limits on decaying dark matter \cite{Ando:2015qda} (based on \cite{Fermi-LAT:2014ryh}). These alternative bounds are found to be comparable to the limits obtained above where comparison can be made. Appendix \ref{A7} shows a comparison to these other limits.

\section{More on Velocity Dependent Annihilations}
\label{S6}

In  previous sections we have defined $p$-wave dark matter to be the case with $\sigma_s=0$ and $\sigma_p\neq0$. Let us now refer to this case as `pure $p$-wave', since while from a model independent perspective this is very clean,  in full models what one really expects is that the $s$-wave channel is non-zero but negligible, such that the $p$-wave annihilation is dominant for all relevant processes. Indeed, arranging for $\sigma_s$ to be zero at all orders in perturbation theory would be very difficult and in cases that the $p$-wave diagram is the dominant annihilation route there are generically $s$-wave annihilation diagrams which exhibit either (multi-level) loop, phase space, and/or chirality flip suppression (see e.g.~\cite{Kumar:2013iva}).

Specifically, to realise $p$-wave annihilating dark matter at freeze-out in a given model one typically requires that the velocity suppression, which is of order $\frac{3}{2x_F} \sim \frac{1}{20}$, is less significant than the relative suppression of the $s$-wave cross section compared to the $p$-wave. A natural example is the case in which $p$-wave annihilation proceeds via tree-level $2\rightarrow2$ diagrams, whereas the $s$-wave annihilations are loop or $2\rightarrow3$ processes, see e.g.~\cite{Bell:2017irk,Chiang:2019zjj}.   One can  consider more complicated scenarios for arriving at relative enhancements of $p$-wave over $s$-wave processes, see for instance \cite{An:2016kie,Kim:2019qix}.

Let us recall the parameterisation of the annihilation cross section
\beq
\langle \sigma v\rangle &=  \sigma_s +\sigma_p  \frac{3}{2x} +\sigma_d  \frac{15}{8x^2}+\cdots 
\eeq
For $p$-wave processes to be dominant at freeze-out one typically requires a  suppression factor 
\beq
\mathcal{F}\equiv \frac{\sigma_s}{\sigma_p}\lesssim\mathcal{O}(10^{-3}),
\eeq
 this is quite achievable if the dominant $s$-wave diagram is loop-induced or exhibits chirality flips. We will refer to scenarios with $\mathcal{F}\neq0$ as `mixed $p$-wave' annihilation, and $\mathcal{F}=0$ corresponds to the pure $p$-wave case.

The important point to recall is that the dark matter velocity in the halo is not the velocity at freeze-out, but will vary within the halo according to (assuming circular orbits) 
\beq\label{6.1}
v(r)=\sqrt{\frac{G}{r}(M_\bullet+M_{\rm halo}(r))} 
\eeq
where $M_{\rm halo}(r)$ is the halo mass out to radius $r$, and thus the velocity follows the total mass enclosed\footnote{We highlight that when stating this velocity dependence earlier in the paper we have made the approximation $M_\bullet+M_{\rm halo}(r)\approx M_\bullet$ which is generally true apart from at large radial distance. This is examined further in Appendix \ref{A1}. In our numerical calculations we use the full relationship, as given in eq.~(\ref{6.1}).} at a given radius: $M_\bullet+M_{\rm halo}(r)$. Therefore, at large radial distances one anticipates that the velocity suppression to $p$-wave processes will be such that $x\gg x_F$ and thus there is a critical radius $r_c$ at which the dominant annihilation processes transitions from $p$-wave to $s$-wave. This is important because the transition alters the inner structure of the halo profile and as a result this impacts the experimental limits on $p$-wave models. Hence while it may be cleaner to simply set  $\sigma_s=0$, more accurate estimates of the experimental bounds may be obtained by assuming $\sigma_s\neq0$ with motivated values for the suppression factor~$\mathcal{F}$.

\begin{figure}[t!]
\centerline{
 \includegraphics[scale=0.44]{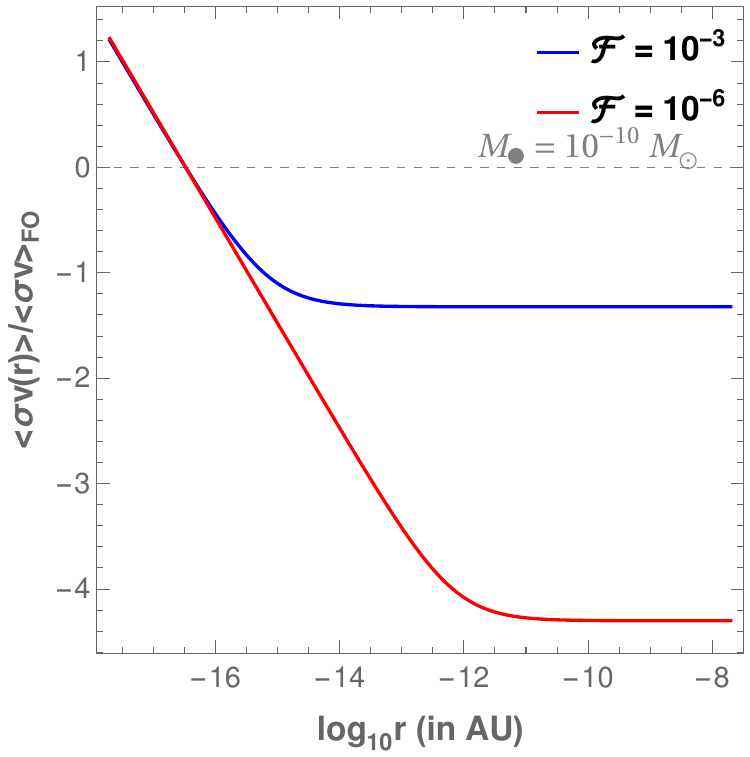}
\hspace{15mm}
 \includegraphics[scale=0.42]{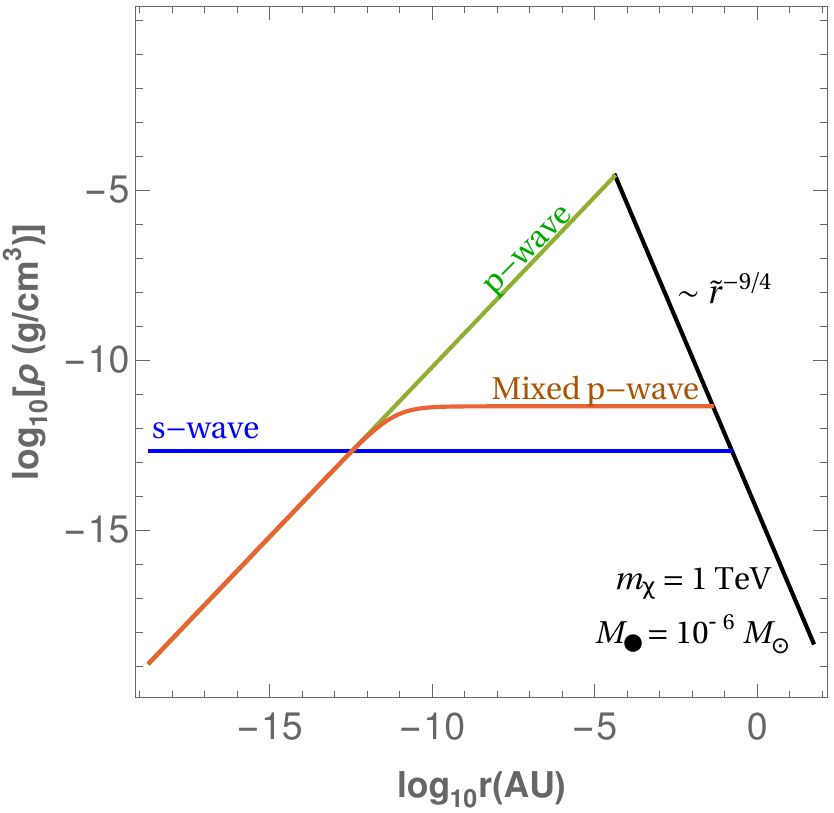}}
\caption{ 
Left.~The dark matter annihilation cross section at radius $r$ compared to the cross section at freeze-out for a PBH of mass $M_\bullet=10^{-10}~M_\oplus$, dark matter mass $m_\chi=$1 TeV (however, the figure is largely insensitive to this value), assuming that the $s$-wave contribution is suppressed by a factor $\mathcal{F}$ relative to the $p$-wave piece. For larger radial distance the dark matter velocities are smaller and thus the suppression to the $p$-wave contribution is larger, at a certain radius the $s$-wave piece eventually dominates and the curve plateaus since this term is velocity independent.  Right.~The dark matter halo profile, similar to Figure \ref{fig:2} (right), here we show the mixed $p$-wave  case taking $\mathcal{F}=10^{-3}$, observe that the mixed case smoothly transitions between the pure $s$-wave and pure $p$-wave cases.  
\label{Fig:svp}}
 \vspace{-2mm}
\end{figure}

In particular, if one assumes a loop factor suppression in the $s$-wave channel then a reasonable estimate for $\sigma_s/\sigma_p$ is of order 
\beq
\mathcal{F}_{\rm loop}\sim g^2/16\pi^2 \sim 10^{-3}\left(\frac{g}{0.5}\right)^2~,
\eeq
where $g$ is some coupling constant of unspecified origin. Taking this unspecified coupling to be $\mathcal{O}(1)$ we arrive at a suppression of order $10^{-3}$. Smaller couplings will allow for greater suppression. Moreover, tree level $s$-wave diagrams involving chirality flip can readily lead to significant suppressions in the cross section, characteristically of the order
\beq
\mathcal{F}_{\rm chiral}\sim \left(\frac{m}{m_\chi}\right)^2 \sim 10^{-6}\ \left(\frac{10^{-3}}{m_\chi/m}\right)^2~,
\eeq
where $m$ is typically the mass of the annihilation product. A modest hierarchy between the two mass scales results in a sizable suppression.
We will take these two suppression factors ($\mathcal{F}_{\rm loop}\sim10^{-3}$ and $\mathcal{F}_{\rm chiral}\sim10^{-6}$) of the $s$-wave relative to the $p$-wave as characteristic. Observe that these suppression factors give the ratio $\sigma_s/\sigma_p$  for $x=1$ and for $x>1$ the $p$-wave develops the velocity suppression thus making the $s$-wave more competitive. Since at freeze-out typically $x_F\sim30$ these $s$-wave suppression factors of $10^{-3}$ and $10^{-6}$ are both sufficient to allow for the relic density of particle dark matter to be set via $p$-wave annihilations.

\begin{figure}[t!]
\centerline{ \includegraphics[scale=0.47]{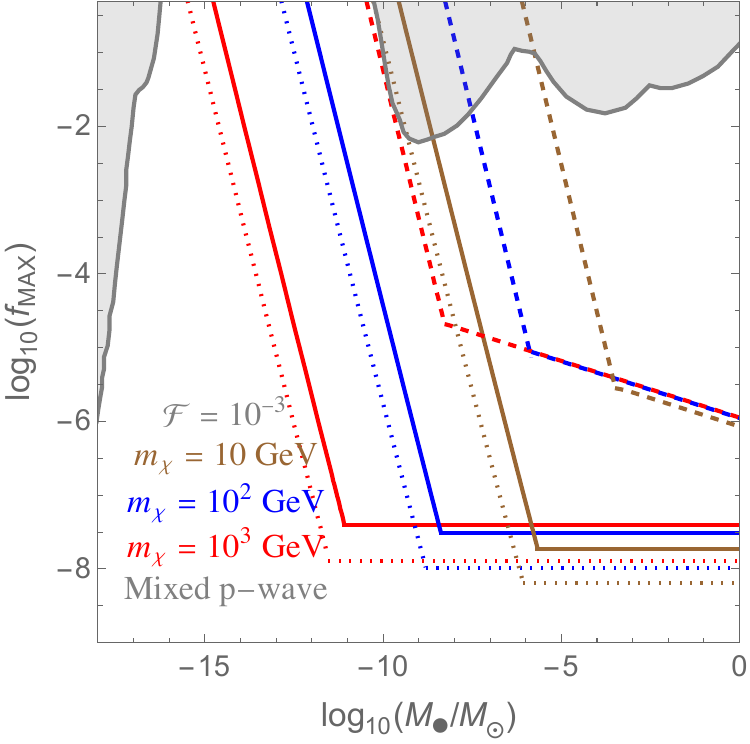}
\hspace{20mm}
 \includegraphics[scale=0.47]{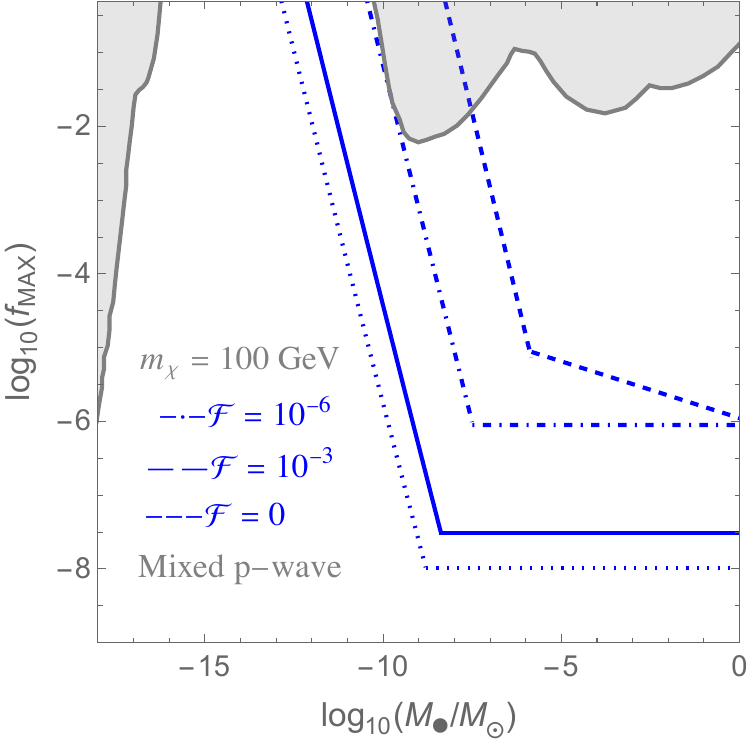}}
\caption{Left.~Extragalactic $\gamma$-ray constraints on mixed $p$-wave annihilating dark matter with $\mathcal{F}=10^{-3}$ (solid curves), analogous to Figure \ref{Fig:sWave-fboundMBH-Galactic}. We show the pure $p$-wave case as the dashed curve, and the pure $s$-wave case dotted. Right.~Similar to the left panel, however here we show how the limits change with the suppression factor $\mathcal{F}$. Note that $\mathcal{F}=0$ corresponds to the pure $p$-wave case.  The inclusion of even highly suppressed $s$-wave interactions is seen to significantly strengthen the bounds.
 \label{Fig:svp2}}
\vspace{-6mm}
\end{figure}

As noted above, at a certain radius $r_c$ the inner structure of the halo switches from being determined by the $p$-wave annihilations to $s$-wave. The value of $r_c$ depends $\sigma_s$, $\sigma_p$, as well as the dark matter mass and black hole mass. The addition of more parameters provides a greater degree of freedom. In Figure \ref{Fig:svp} (left panel) we compare the $s$-wave annihilation cross section to the $p$-wave cross section (for two values of $\mathcal{F}$) as a function of radial distance from the PBH taking the dark matter mass to be $m_\chi=1$ TeV and the PBH mass to be $M_\bullet=10^{-10}~M_\odot$.  We highlight that  the central region is dominated by the $p$-wave process and then transitions to $s$-wave domination for $r\gtrsim r_c\simeq 10^{-16}$ AU (with the stated parameters).
To provide further intuition in  Figure \ref{Fig:svp} (right) we show the halo profiles for $\mathcal{F}\sim10^{-3}$, we assume the same dark matter mass and PBH mass as in the left panel. Correspondingly, in  Figure \ref{Fig:svp2} we derive the experimental limits for $\mathcal{F}\sim10^{-3}$ and $\mathcal{F}\sim10^{-6}$, this is analogous to Figure \ref{Fig:sWave-fboundMBH-Galactic} but where we no longer assume that $\sigma_s=0$, but rather $\sigma_s=\mathcal{F}\sigma_p$. Observe that the mixed $p$-wave case provides a transition between the pure $s$-wave and pure $p$-wave cases.

Finally, we highlight that these considerations are also a factor for $d$-wave dark matter, and the velocity dependence in the halo implies that the true halo structure can exhibit radial shells corresponding to transitions between a $d$-wave region, followed by a $p$-wave shell and then an $s$-wave plateau, prior to the transition to the halo profiles of Section \ref{S2}. Indeed, in the $d$-wave  model outlined in Appendix \ref{A3} the $d$-wave suppressed tree level diagrams are accompanied by velocity independent $2\rightarrow3$ diagrams which are phase space suppressed. Since $d$-wave dark matter occurs far less frequently as motivated models compared to $p$-wave scenarios, we will not explore this case in further detail here.

\section{Concluding Remarks}
\label{S7}

Primordial black holes could well exist as remnants of early universe cosmology but not be sufficiently abundant to account for the anomalous gravitational observations that have led to the inference of dark matter.  In such a case it is natural to suppose that PBH could co-exist with a cosmological abundance of dark matter particles and, indeed, in scenarios of physics beyond the Standard Model potential dark matter candidates are ubiquitous.

We have highlighted that in the case that PBH and particle dark matter co-exist in appreciable abundances then the limits from indirect detection can be very constraining in the case that the relic density of particle dark matter is set via thermal freeze-out. Notably, our work furthers earlier studies in the literature by using a more careful treatment of the PBH halo profile and simultaneously extending this analysis away from the velocity independent case to the richer scenarios involving $p$-wave and $d$-wave dark matter. 

Our main findings are illustrated by Figures \ref{Fig:sWave-fboundMBH-Galactic} \& \ref{Fig:svp2}, we highlight that velocity dependent dark matter annihilations ameliorate the bounds from indirect detection both in the sense that these models allow for significantly larger values of $f_{\rm PBH}$ to be consistent with observational constraints. Thus while in the case of $s$-wave dark matter one require that the PBH fraction be sub-1\% in order to avoid constraints for $M_\bullet\gtrsim10^{-10\pm2}M_\odot$ (assuming Weak scale dark matter of mass $m_\chi\gtrsim10^{2\mp1}$ GeV) for (pure) $p$-wave dark matter this bound weakens to $M_\bullet\gtrsim10^{-7\pm2}M_\odot$, and for $d$-wave models even heavier PBH are permitted with a bound $M_\bullet\gtrsim10^{-5\pm2}M_\odot$. We highlight that these $p$ and $d$-wave bounds assume that the cross sections with less velocity suppression are exactly vanishing and, as discussed in Section \ref{S6} a more realistic analysis leads to stronger constraints, see Figure \ref{Fig:svp} (right). 

The limits derived above assume that the entire PBH population have the same mass and negligible spin. This monochromatic spectrum assumption provides for reasonable estimates (which is the aim of this work), however, it is plausible  that the PBH would have extended mass distribution functions, such as lognormal or powerlaw spectra, see e.g.~\cite{Carr:1975qj,Carr:2017jsz,Kuhnel:2017pwq}. Moreover, while it is thought that PBH formed during radiation domination will typically be mostly slowly rotating \cite{Chiba:2017rvs,Mirbabayi:2019uph} away from this scenario PBH can exhibit appreciable (even extreme) spins, see e.g.~\cite{Harada:2017fjm,Eroshenko:2021sez,Dvali:2021byy}. 
 We intend to explore the bounds on the mixed PBH-particle dark matter  scenario in these more complicated settings in subsequent work.

\vspace{4mm}
\noindent
{\bf Acknowledgments.}
JS is supported by the Departments of Excellence grant awarded by the Italian Ministry of Education, University and Research (Miur), and research grant `The Dark Universe: A Synergic Multimessenger Approach', No.~2017X7X85K funded by the Italian Ministry of Education, University and Research (Miur), and Research grant `TAsP (Theoretical Astroparticle Physics)' funded by Instituto Nazionale di Fisica Nucleare.
JU is supported by NSF grant PHY-2209998 and is grateful for the hospitality of the Berkeley Center for Theoretical Physics during the completion of this work.
We thank Estanis Utrilla Gines, Luca Visinelli, and Sam Witte for helpful discussions.

\appendix
\section{Appendices}

\subsection{Mass Accreted by the Black Hole}
\label{A1}

In this Appendix we derive an analytic expression for the accreted dark matter mass in the halo by a black hole and investigate when the halo mass becomes important while studying the orbital motion of the dark matter particles. For simplicity, we assume $s$-wave dark matter.

\begin{figure}[t!]
\centerline{
 \includegraphics[scale=0.44]{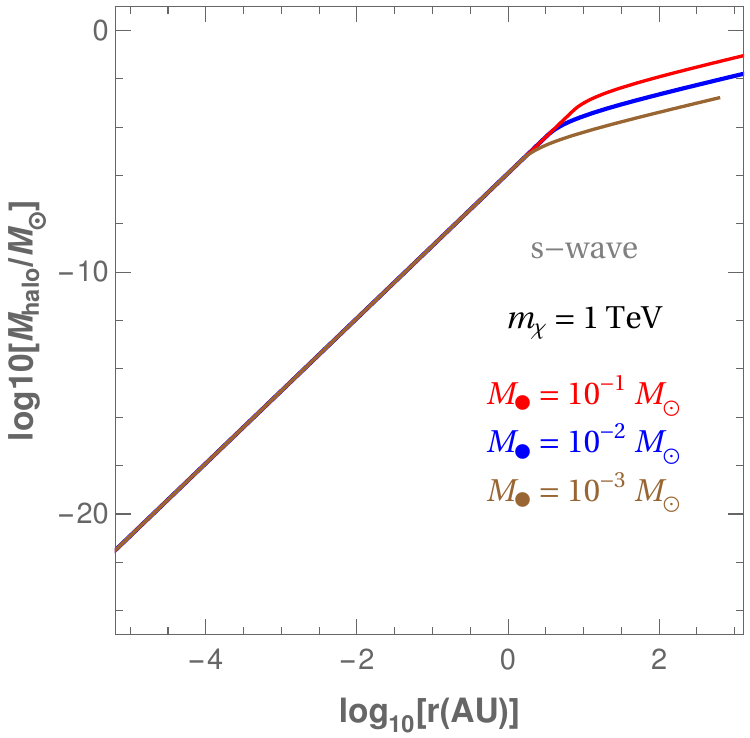}
 \includegraphics[scale=0.44]{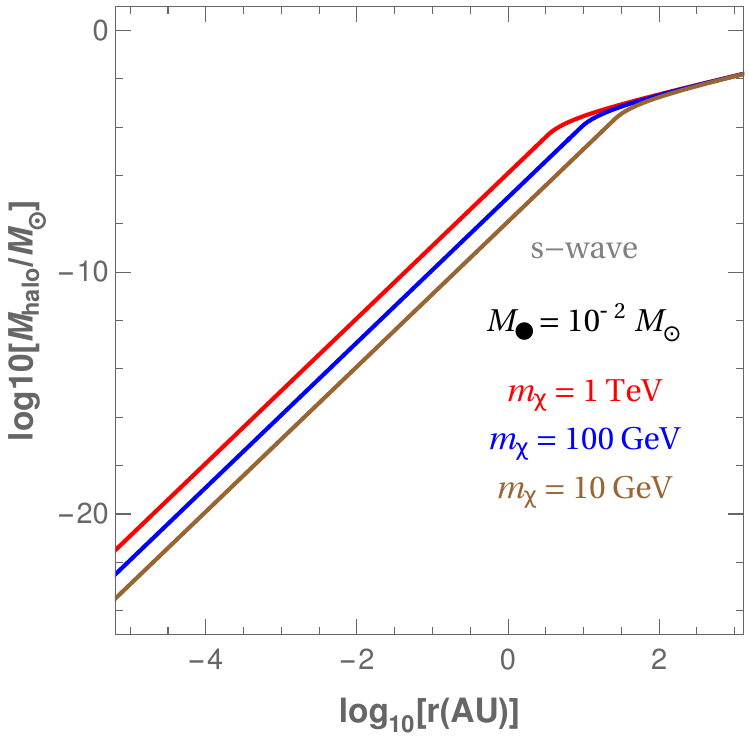}
 \includegraphics[scale=0.44]{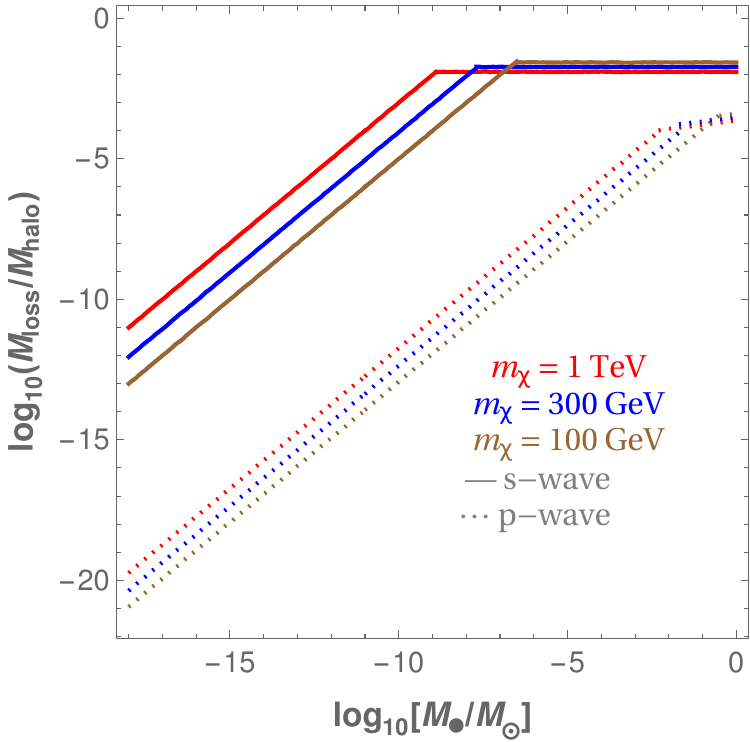}}
\caption{\label{fig6}
Left/Centre: Mass of the dark matter halo contained within radial distance $r$, denoted $M_{\rm halo}$ for fixed choice of either dark matter mass (left) or PBH mass (centre), as the other mass parameter is varied. This is presented for $s$-wave annihilation cross section and thus assumes a central plateau in the density profile. Note that the mass of $M_{\rm halo}<M_{\bullet}$ out to  $\mathcal{O}(1)$~AU for the cases presented. Right.~The fraction of dark matter removed from the PBH halo through annihilations $M_{\rm loss}/M_{\rm acc}\equiv(M_{\rm acc}-M_{\rm halo})/{M_{\rm acc}}$ this is shown as a function of PBH mass $M_\bullet$. The  $s$-wave (pure $p$-wave) case is shown solid (dotted).}
\end{figure}

Neglecting initially dark matter self-interactions,  the mass of the dark matter halo $M_{\rm acc}$ (the total mass of the accreted dark matter prior to annihilations), can be found for  heavy PBH ($M_{\bullet}>M_{2}$) by integrating over the 9/4th profile 
\beq \label{ea1}
\frac{M_{\rm acc}}{M_{\bullet}} &= \frac{ G^{3}M^{2}_{\bullet}}{3c^{6}}\frac{\left(128\pi\right)^{3/2}}{\Gamma^{2}\left(1/4\right)}\rho_{i}^{\rm kd}\tilde{r}_{\rm kd}^{9/4}\tilde{r}_{\rm eq}^{3/4}\left[1-\frac{1}{6\left(2\pi^{3}\right)^{1/2}}\right].
\eeq

In contrast when including dark matter annihilations the halo mass is determined by integrating over the annihilation plateau near the center, and the 9/4  profile at larger radial distances. Thus for heavy PBH ($M_{\bullet}>M_{2}$), the total mass in the halo following depletion via annihilations is given by
\beq
\frac{M_{\rm halo}}{M_{\bullet}}& = \frac{32\pi G^{3}M^{2}_{\bullet}}{c^{6}} \left[\frac{m_{\chi}}{3\langle\sigma v\rangle t_{\rm age}}\tilde{r}_{\rm core}^{3}+\frac{\sqrt{128\pi}}{\Gamma^{2}\left(1/4\right)}\rho_{i}^{\rm kd}\tilde{r}_{\rm kd}^{9/4}\left[\frac{4}{3}\left(\tilde{r}_{\rm eq}^{3/4}-\tilde{r}_{\rm core}^{3/4}\right)-\frac{2r_{\rm eq}^{-3/4}}{9\left(2\pi^{3}\right)^{1/2}}\left(\tilde{r}_{\rm eq}^{3/2}-\tilde{r}_{\rm core}^{3/2}\right)\right]\right].
\eeq

If Figure \ref{fig6} (left) we show how the total mass of the dark matter halo (including the effects of dark matter annihilations) enclosed at a radius $r$ varies with PBH mass for a fixed choice of dark matter mass. We also show (centre panel) how the halo mass enclosed at radius $r$ varies with dark matter mass for a fixed PBH mass. Finally, in the right panel we present the relative change in the total mass of the halo after depletion through self-annihilations, being the fractional change $M_{\rm loss}/M_{\rm acc}\equiv(M_{\rm acc}-M_{\rm halo})/{M_{\rm acc}}$, as a function of PBH mass.  While in eq.~(\ref{ea1}) we give the total mass assuming a 9/4 profile, in Figure \ref{fig6} (right) we use the full PBH halo profiles outlined in Section \ref{S2}. 

In particular, we highlight that characteristically $M_{\rm acc}\sim M_\bullet$, hence it follows that the PBH dominates the gravitational force out to large radii and the dark matter velocity approximation $v(r)\simeq\sqrt{GM_\bullet/r}$ (which neglects the mass contribution due to the halo) is robust.
Additionally, we highlight that the fractional change of the total mass due to annihilations is typically sub-percent level, $M_{\rm loss}/M_{\rm acc}\lesssim0.01$, and thus we do not anticipate subsequent significant rearrangements of the halo profile following depletion due to annihilations. 

\subsection{PBH stripping by PBH}
\label{Anew}

The prospect of PBH halo stripping due to encounters with other PBH is complicated due to the unknown abundances and distribution of PBH, but also due to the fact that unlike stars the second PBH has a dark matter halo of its own. For a rudimentary estimate, we make a similar argument to the stellar stripping calculation in Section \ref{3.4}. Since $M_{\rm halo}\sim M_{\rm PBH}$ we expect the stripping radius to be
\beq
r_T[{\rm PBH}]\sim d\left(\frac{M_{\rm halo}}{2M_{\rm PBH}}\right)^{1/3}\sim d
\eeq
 where $d$ is the distance of closest approach between a specific PBH and another PBH in the galactic population.
Given the total mass of dark matter in the Milky Way is $\sim10^{12}M_\odot$ the number of PBH in the Milky Way is 
\beq
N_{\rm PBH}\sim f_{\rm PBH}10^{12}M_\odot/M_{\rm PBH}.	
\eeq
For simplicity let us assume that the PBH are uniformly spaced on a regular square lattice with each side being the diameter of the Milky Way halo $\sim500$ kpc. Thus the horizontal/vertical spacing of adjacent lattice sites is
\beq
l=\frac{D}{N_{\rm PBH}}\sim 5\times10^{-4} {\rm pc}\left(\frac{10^{-3}}{f_{\rm PBH}}\right)\left(\frac{M_{\rm PBH}}{M_\odot}\right)~.
\eeq
Similar to Section \ref{3.4} we estimate the separation during the closest encounter for a specific PBH traversing this lattice to be
\beq
d=\frac{l}{\sqrt{N_{\rm PBH}}}=\frac{D}{N_{\rm PBH}^{3/2}}=10^{-8}~{\rm pc}\left(\frac{10^{-3}}{f_{\rm PBH}}\right)^{3/2}\left(\frac{M_{\rm PBH}}{M_\odot}\right)^{3/2}~.
\eeq
For solar mass PBH and taking $f_{\rm PBH}$ near the exclusion bounds, we find $r_T$ is a factor of 100 smaller than the stellar stripping found in Section \ref{3.4}. However, for smaller $f_{\rm PBH}$, PBH stripping will be less important than stellar stripping. 
For lighter PBH, say $10^{-6} M_\odot$ PBH, the encounter rate goes up and thus the stripping radius goes down:
\beq
 r_T \sim d\sim 10^{-17}~{\rm pc}\left(\frac{10^{-3}}{f_{\rm PBH}}\right)^{3/2}\left(\frac{M_{\rm PBH}}{10^{-6}M_\odot}\right)^{3/2}		
\eeq
However, the schwarzschild radius for a $10^{-6} M_\odot$ PBH is  $\sim10^{-13}$pc, suggesting that, in fact, the dark matter halo may be entirely disrupted during this encounter.  However, unlike close encounters with stars, close encounters with other PBH implies passing through the dark matter halo around the second PBH. Thus, the halos of both PBHs are disrupted. It is not entirely clear whether they will be entirely dissipated through the close encounter or if some fraction will remain. Regardless, one would also expect the profiles to be strongly perturbed. 

The scaling estimate above is rather crude and a simulation of the galactic dynamics using realistic PBH distributions should be used to address this question. Thankfully, such  PBH-PBH encounters should not significantly impact the extragalactic bounds studied here since we expect them to mostly occur after galaxy formation $z\sim\mathcal{O}(1)$, but this may be important when calculating bounds on galactic PBH populations.

\subsection{Dark Matter Self-Scattering}

Dark matter self-scattering is typically important for dark matter models with large self-interactions introduced through new states that couple only to the dark matter. However, in the case of PBH halos one may be concerned that the densities are sufficiently high that, even in the absence of large tree level dark matter self-interactions,  loops of Standard Model states can induce non-negligible  dark matter self-scattering. Indeed, since the dark matter relic density is assumed to be set via freeze-out these loop induced interactions are irremovable. Such dark matter self-scattering allows for energy transport within the halo, which can alter the density profile. This process is highly analogous to self-interacting dark matter solutions to the core-cusp problem, see e.g.~\cite{Spergel:1999mh}. Notably, if the cross section required to set the correct relic density of particle dark matter would imply large self-interactions, then the profile derived in Section \ref{4.2} may not be appropriate.

Next, we comment on how self-interactions of the dark matter  halo around the PBH could significantly change the overall picture. In principle the dark matter  annihilation into Standard Model states always induces a self-interaction diagram through loop processes. However, the loop-induced self-interaction matrix elements typically represent a lower bound on the self-interactions in the dark matter  sector. Furthermore, the precise way the loop-induced process generate the self-interactions is dependent on the full hidden sector model.

As an example, consider the operator responsible for dark matter  annihilation into $b$ quarks: $\frac{1}{\Lambda^2}\bar{\psi} \psi \bar{b}b$. The loop-induced self-interaction from a loop of $b$-quarks is UV-completion dependent, since the $b$-quark loop is quadratically divergent. Suppose that the scattering between the two sectors in the UV is realized by $Z'$ mixing with the $B$ boson. Then the $\bar{\chi} \chi \to \bar{\chi} \chi$ process has a parametric dependence of form:
\begin{equation}
\sigma_{\rm SIDM} \propto g^4 M_{\chi}^2 /M_{Z'}^4 \sim g^4/\Lambda^2,
\end{equation}
where for simplicity we assume the dark sector has a typical shared scale  $\Lambda\sim M_{\chi}\sim M_{Z'}$, which would be natural if the masses for both $\chi$ and $Z'$ arise from some dark sector Higgs mechanism. The annihilation into Standard Model states has a cross-section of parametric form:
\begin{equation}
\sigma_{A} \propto \frac{g_2^2 g^2 \sin^2\theta M_{\chi}^2 }{M_{Z'}^4} \sim \frac{g_2^2 g^2 \sin^2\theta}{\Lambda^2},
\end{equation}
where  $\theta$ represents the $Z'$-$B$ mixing angle, and $g_2$ is the Standard Model weak coupling.

\begin{figure}
\centerline{ \includegraphics[width=0.48\textwidth]{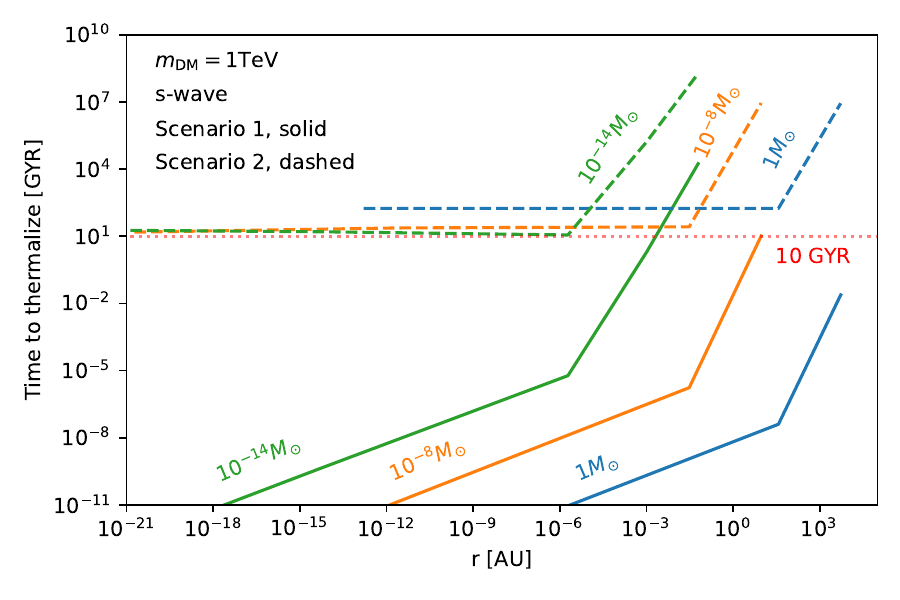}~~~~
 \includegraphics[width=0.48\textwidth]{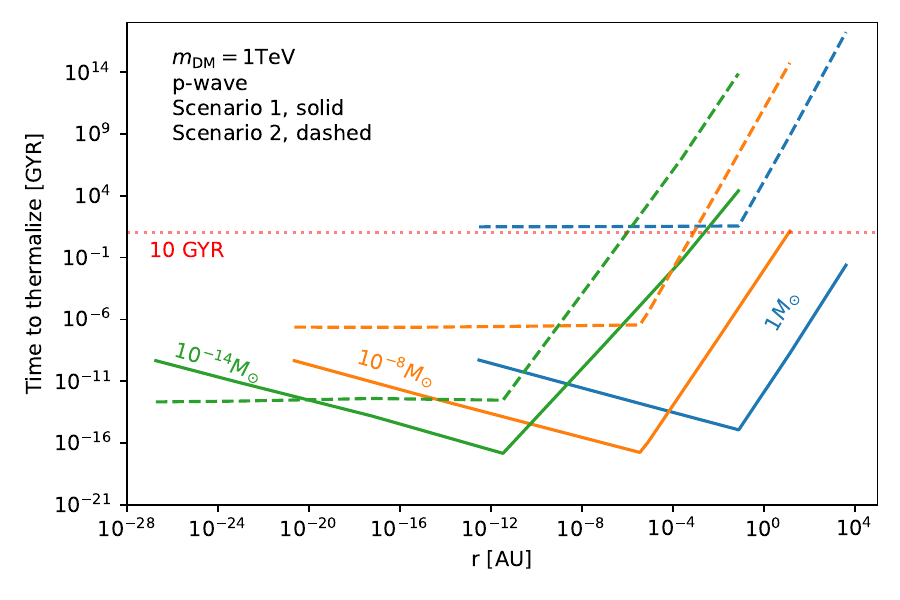}}
\vspace{-3mm}
\caption{Thermalization times inside the dark matter halo around the PBH as a function of the distance from the PBH. In both plots we set the dark matter mass to $1$ TeV, and probe three different PBH masses: $1M_{\odot},~10^{-8} M_{\odot}$ and $10^{-14} M_\odot$ denoted by blue, orange and green colors, respectively. The case of the $s$-wave scattering is on the left, the case of $p$-wave thermal dark matter is on the right. 
The solid lines correspond to Scenario 1 from the text (the self interaction cross-section is set by the current limit of $\sigma/m = 0.1 \mathrm{g}/\mathrm{cm}^2$) , while Scenario 2 is achieved by setting the self-scattering cross-section to be the same as the thermal cross-section with the Standard Model particles.\label{Fig7}}
\end{figure}

Parametrically, the ratio of the two is given by $(g_2 \sin(\theta)/g)^2$, possibly not far from unity (but highly dependent on parameter choices). Provided that the separation of the two sectors is achieved through low mixing, then the self-interactions should lead to larger cross-sections.
Since there is a great deal of model freedom, we are going to consider here just two scenarios:

\begin{itemize}
\item In Scenario 1, we take the self-interaction cross-section to be the current limit for dark matter self interactions  from astrophysical observation: $\sigma_{\rm SIDM} = 0.1 \mathrm{cm}^2/\mathrm{g}$ \cite{Tulin:2017ara,Bullock:2017xww}.
\item In Scenario 2, we rather set the self-interaction cross-section to be the same as the $s$-wave thermal freeze-out cross-section: $\sigma_{\rm SIDM}=\langle \sigma v \rangle = 2\times 10^{-26} \mathrm{cm^3/s}$.
\end{itemize}
The first represents an upper bound, the second represents a reasonable benchmark scattering rate, however there is no requirement that the scattering and annihilation cross sections be similar and thus this could lead to an under or over estimate. Regardless, without specifying a detailed model Scenario 2 provides a reasonable benchmark to compare to. Moreover, contrasting between the two scenarios will lead to an interesting observation.\\

\noindent {\bf Scenario 1.}~The time for dark matter  to thermalize inside the dark matter  halo around the PBH is given by:
\begin{equation}
 t_{\rm th} = \mathcal{O}(1) \left(\sigma_{\rm SIDM} n(r) v(r)\right)^{-1}~.
\end{equation}
We take the dark matter  density profiles of Section \ref{S2} \& \ref{S4},  and the assumption $v^2(r) \sim GM_{\bullet}/r$. This allows us to evaluate the thermalization times as a function of distance from the PBH, $r$. We show these results in Figure \ref{Fig7}  for $s$-wave and $p$-wave models for two different dark matter  masses: $100$~GeV and 1~TeV and for three PBH masses $[10^{-14}, 10^{-8},1]\times M_{\odot}$. 

As we can see, for Scenario 1, the thermalization times are vastly shorter than the existence of these systems. In the $s$-wave models for which the density profile is very close to isothermal anyway (flat and then $r^{-9/4} \sim r^{-2}$) this may not have strong consequences. However, in $p$-wave models this will lead to thermalization, which will flatten the halo inner density profile, likely transport more dark matter into the inner profile and increase the annihilation rate, until there is no more material to transport further in. As a result, we would expect that in these models, the extragalactic bounds (which are dependent on early annihilation rates) will become stronger, while the galactic bounds which are dependent on current density profiles, will correspondingly weaken.

\vspace{5mm}

\noindent {\bf Scenario 2.} We can see that for the solar mass PBH the thermalization time begins to approach the age of the Universe and for lighter black holes or larger cross-sections the effect of transporting dark matter becomes relevant and leads to increased historical annihilation rates. A careful analysis of the scenario obviously requires more work and we plan to explore this in an upcoming dedicated paper.


\subsection{$d$-wave Dark Matter}
\label{A3}

In this Appendix we sketch a model of $d$-wave dark matter that has been previously identified in the literature. While there are many well known examples of $s$-wave and $p$-wave annihilating dark matter (see e.g.~\cite{Shelton:2015aqa} for examples of the latter), we include here an example of the less common case in which $d$-wave annihilations are the dominant process, highlighting the $d$-wave model of \cite{Giacchino:2013bta}, later discussed in \cite{Boddy:2019qak}.
Consider the case of real scalar dark matter $\chi$  that interacts with the Standard Model through heavy vector-like fermions and annihilates into lepton-antilepton pairs. Specifically, the dark matter interacts with right-handed Standard Model leptons, $l_{R}$, through heavy vector-like leptons $\Psi$ which are SU$(2)_{L}$ singlet, such that the hypercharges for $\Psi$, and $l_{R}$ are the same, $Y_{\Psi} = Y_{l_{R}}$, with the Lagrangian contribution 
\beq\label{eq:d-wave-lagrangian}
\mathcal{L}\supset y\chi \bar{\Psi}l_{R}+{\rm h.c.}\cdots,
\eeq
where $\chi$ is the real scalar dark matter candidate, $y$ is a Yukawa coupling. In the interaction above, we assume a discrete $Z_{2}$ symmetry $S\to -S$ and $\Psi\to -\Psi$. The lightest member will be stable under this $Z_{2}$ symmetry, which we take to be $\chi$, such that $m_{\chi}<m_{\Psi}$.

\begin{figure}[t]
\begin{center}
\begin{tikzpicture}[node distance = 2 cm and 2.5 cm]
\coordinate[label = left:$\chi$](e1);
\coordinate[below = 2 cm of e1 , label = left:$\chi$](e2);
\coordinate[right = 2.0 cm of e1](aux1);
\coordinate[right = 2.0 cm of e2](aux2);
\coordinate[right = 4 cm of e1, label = right:$l$](e3);
\coordinate[right = 4 cm of e2, label = right:$\bar{l}$](e4);

\draw[scalar] (e1) -- node[]{} (aux1);
\draw[scalar] (e2) -- node[]{} (aux2);
\draw[fermion] (aux2) -- node[]{} (aux1);
\draw[fermion] (e4) -- node[]{} (aux2);
\draw[fermion] (aux1) -- node[]{} (e3);
\end{tikzpicture}
~~~~~~
\begin{tikzpicture}[node distance = 2 cm and 2.5 cm]
\coordinate[label = left:$\chi$](e1);
\coordinate[below = 2 cm of e1 , label = left:$\chi$](e2);
\coordinate[right = 2.0 cm of e1](aux1);
\coordinate[right = 2.0 cm of e2](aux2);
\coordinate[right = 4 cm of e1, label = right:$l$](e3);
\coordinate[right = 4 cm of e2, label = right:$\bar{l}$](e4);

\draw[scalar] (e1) -- node[]{} (aux2);
\draw[scalar] (e2) -- node[]{} (aux1);
\draw[fermion] (aux2) -- node[]{} (aux1);
\draw[fermion] (e4) -- node[]{} (aux2);
\draw[fermion] (aux1) -- node[]{} (e3);
\end{tikzpicture}

\end{center}
\caption{Two-body annihilation of the real scalar dark matter $\chi$ to Standard Model leptons $l$ mediated by heavy vector-like fermion via $t$ and $u$ channels. These diagrams are  $d$-wave suppressed.}
\label{Fig:ScalarDM-heavyVectorFermion-2body}
\end{figure}
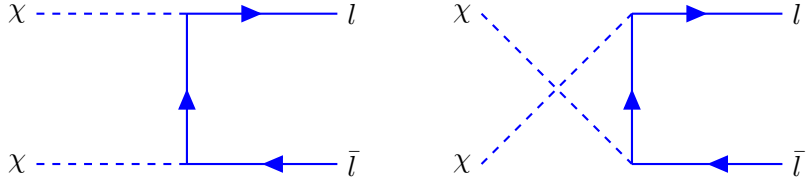

In the chiral limit $m_{f}\to 0$ the thermally averaged cross-section for the two-body annihilation process $\chi\chi\to l\bar{l}$ (which is $t$ and $u$-channel, as shown in Figure \ref{Fig:ScalarDM-heavyVectorFermion-2body}) is given by 
\beq\label{eq:sigma-2body}
\langle\sigma v\rangle\approx\frac{y^{4}}{60\pi}\frac{v^{4}}{m_{\chi}^{2}}\frac{1}{(1+z^{2})^{4}},
\eeq
at the leading order in $v$, with $z = m_{\Psi}/m_{\chi}$. 

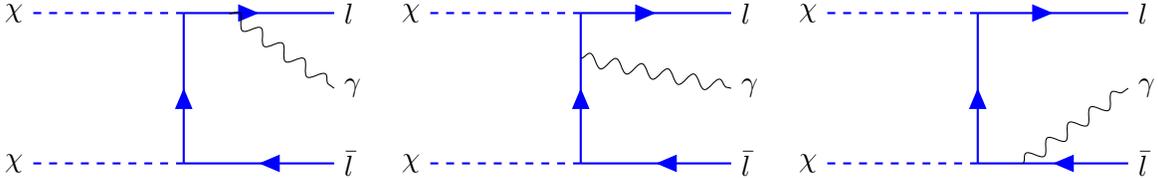
\begin{figure}[t]
\begin{tikzpicture}[node distance = 2 cm and 2.5 cm]
\coordinate[label = left:$\chi$](e1);
\coordinate[below = 2 cm of e1 , label = left:$\chi$](e2);
\coordinate[right = 2.0 cm of e1](aux1);
\coordinate[right = 2.0 cm of e2](aux2);
\coordinate[right = 4 cm of e1, label = right:$l$](e3);
\coordinate[right = 4 cm of e2, label = right:$\bar{l}$](e4);
\coordinate[right = 0.6 cm of aux1](aux3);
\coordinate[above= 1.0 cm of e4, label = right:$\gamma$](e5);

\draw[scalar] (e1) -- node[]{} (aux1);
\draw[scalar] (e2) -- node[]{} (aux2);
\draw[fermion] (aux2) -- node[]{} (aux1);
\draw[fermion] (e4) -- node[]{} (aux2);
\draw[fermion] (aux1) -- node[]{} (e3);
\draw[photon] (aux3) -- node[]{} (e5);
\end{tikzpicture}
~
\begin{tikzpicture}[node distance = 2 cm and 2.5 cm]
\coordinate[label = left:$\chi$](e1);
\coordinate[below = 2 cm of e1 , label = left:$\chi$](e2);
\coordinate[right = 2.0 cm of e1](aux1);
\coordinate[right = 2.0 cm of e2](aux2);
\coordinate[right = 4 cm of e1, label = right:$l$](e3);
\coordinate[right = 4 cm of e2, label = right:$\bar{l}$](e4);
\coordinate[below = 0.6 cm of aux1](aux3);
\coordinate[above= 1.0 cm of e4, label = right:$\gamma$](e5);

\draw[scalar] (e1) -- node[]{} (aux1);
\draw[scalar] (e2) -- node[]{} (aux2);
\draw[fermion] (aux2) -- node[]{} (aux1);
\draw[fermion] (e4) -- node[]{} (aux2);
\draw[fermion] (aux1) -- node[]{} (e3);
\draw[photon] (aux3) -- node[]{} (e5);
\end{tikzpicture}
~
\begin{tikzpicture}[node distance = 2 cm and 2.5 cm]
\coordinate[label = left:$\chi$](e1);
\coordinate[below = 2 cm of e1 , label = left:$\chi$](e2);
\coordinate[right = 2.0 cm of e1](aux1);
\coordinate[right = 2.0 cm of e2](aux2);
\coordinate[right = 4 cm of e1, label = right:$l$](e3);
\coordinate[right = 4 cm of e2, label = right:$\bar{l}$](e4);
\coordinate[right = 0.6 cm of aux2](aux3);
\coordinate[above= 1.0 cm of e4, label = right:$\gamma$](e5);

\draw[scalar] (e1) -- node[]{} (aux1);
\draw[scalar] (e2) -- node[]{} (aux2);
\draw[fermion] (aux2) -- node[]{} (aux1);
\draw[fermion] (e4) -- node[]{} (aux2);
\draw[fermion] (aux1) -- node[]{} (e3);
\draw[photon] (aux3) -- node[]{} (e5);
\end{tikzpicture}
\caption{Velocity independent three-body annihilation via Virtual Internal Bremsstrahlung (VIB).}
\label{Fig:ScalarDM-heavyVectorFermion-3body-VIB}
\end{figure}

There is also a competing $2\rightarrow3$ annihilation process $\chi\chi\to \gamma l\bar{l}$ (see Figure \ref{Fig:ScalarDM-heavyVectorFermion-3body-VIB}) for which the thermally averaged cross section is given by \cite{Giacchino:2013bta}
\beq\label{eq:sigma3bodyVIB}
\langle\sigma v\rangle_{\chi\chi\to\gamma l\bar{l}} = \frac{\alpha y^{4}}{4\pi^{2}m_{\chi^{2}}}F(z),
\eeq
where $F(z)$ is defined in terms of Polylogarithms as follows
\beq
F(z) &= \frac{1}{8(z^{2}+z^{4})}\left( 12z^{2}+16z^{4}+\left(-3-13.77z^{2}-z^{4}-1.32z^{6}-8\left(z+z^{3}\right)^{2}\log(z)\right)\log\left(z^{2}-1\right)\right.\\
&\quad\quad + \left(1+5z^{2}+0.54z^{4}-3.45z^{6}+8\log(z)\log(z^{2}+1)\left(z+z^{3}\right)^{2}\right)\\
&\quad\quad\left. + \left(1+z^{2}\right)^{3}\log\left(z^{4}-1\right)+4\left(z+z^{3}\right)^{2}\left({\rm PolyLog}\left[2,\frac{1}{2}-\frac{1}{2z^{2}}\right]-{\rm PolyLog}\left[2,\frac{1}{2}+\frac{1}{2z^{2}}\right]\right)\right).
\eeq
The above expression requires $z>1$, which is satisfied given the mass ordering $m_{\Psi}>m_{\chi}$.

Comparing the $d$-wave suppressed two-body annihilations (eq.~\eqref{eq:sigma-2body}), and three-body VIB annihilations (eq.~\eqref{eq:sigma3bodyVIB}) one finds  for $m_{\Psi}\gg m_{\chi}$ that the $d$-wave process is dominant at freeze-out. Moreover, for $y\sim1$ one finds that the dark matter relic density can be correctly reproduced for $m_\chi\sim100$ GeV.
We refer the reader to \cite{Giacchino:2013bta} for further details of this model and the associated phenomenology.

\subsection{Annihilation rate inside the dark matter  halo}\label{A4}

We next derive analytic forms for the annihilation rate of the dark matter particles inside the halo at different redshifts, $z$, using the density profiles as discussed in Section \ref{S2}. At various $z$, the Hubble parameter is defined as 
\beq
\frac{H(z)}{H_{0}}=\left[\Omega_{\Lambda}+\Omega_{\rm mat}(1+z)^{3}+\Omega_{\rm rad}(1+z)^{4}\right]^{1/2},
\eeq
 where $H_{0}$ is the present day Hubble constant, $\Omega_{\Lambda}$ is the dark energy density, $\Omega_{\rm mat}$ is the matter density, and $\Omega_{\rm rad}$ is the radiation density as observed today. 
 
 For the dark matter mass range in this study the density profile for the heavy mass PBH (with $M_{\bullet}>M_{2}$)  is described by the maximum density core followed by the 9/4 profile, while the 3/2  profile becomes only important for the light mass PBH ($M_{\bullet}<M_{1}$) and intermediate mass  PBH ($M_{1}\leq M_{\bullet}\leq M_{2}$). The 3/4  profile mostly annihilates away into the maximum  density core and remains negligible for light PBH. Thus, it suffices to define the characteristic radial distance at which the 3/2 profile for the light mass and intermediate mass black holes meet the 9/4 profile, given by
  \beq
  \tilde{r}_{B}^{\prime\prime} = \left(\frac{8\pi^{2}}{\Gamma^{2}(1/4)}\right)^{4/3}\tilde{r}_{\rm kd}^{3}x_{\rm kd}^{-2}.
  \eeq

  The radius $r_{\rm core}$, below which dark matter annihilations set the profile, is  sensitive to the PBH mass and whether the dark matter is $s$-, $p$-, or, $d$-wave. For the case that the core region transitions into an exterior profile with $\rho_{\rm i, kd}^{2/3}$ the form of the core radii is given by
\beq\label{A12}
\tilde{r}_{\rm core, 3/2} = \begin{cases} \left(\frac{2}{\pi^{3}}\right)^{1/3}\rho_{\rm i, kd}^{2/3} x_{\rm kd} \frac{\sigma_{s}^{2/3}}{m_{\chi}^{2/3}}H^{-2/3}(z) & \qquad s{\rm -wave} \\[5pt]
 \left(\frac{1}{2\pi^{3}}\right)^{1/5}\rho_{\rm i, kd}^{2/5} x^{3/5}_{\rm kd} \frac{\sigma_{p}^{2/5}}{m_{\chi}^{2/5}}H^{-2/5}(z) & \qquad p{\rm -wave} \\[5pt]
 \left(\frac{1}{8\pi^{3}}\right)^{1/7}\rho_{\rm i, kd}^{2/7} x^{3/7}_{\rm kd} \frac{\sigma_{d}^{2/7}}{m_{\chi}^{2/7}}H^{-2/7}(z) & \qquad d{\rm -wave} 
  \end{cases}~.
\eeq
Alternatively, if the central core transitions into a profile with $\rho_{\rm i, kd}^{9/4}$ the core radius is rather
\beq\label{A13}
\tilde{r}_{\rm core, 9/4} = \begin{cases}  \frac{(128\pi)^{2/9}}{\Gamma^{8/9}(1/4)} \rho_{\rm i,kd}^{4/9}\tilde{r}_{\rm kd}\left(\frac{\sigma_{s}}{m_{\chi}}\right)^{4/9}H^{-4/9}(z)
& \qquad s{\rm -wave} \\[5pt]
\frac{(32\pi)^{2/13}}{\Gamma^{8/13}(1/4)} \rho_{\rm i,kd}^{4/13}\tilde{r}^{9/13}_{\rm kd}\left(\frac{\sigma_{p}}{m_{\chi}}\right)^{4/13}H^{-4/13}(z)
& \qquad p{\rm -wave} \\[5pt]
 \frac{(8\pi)^{2/17}}{\Gamma^{8/17}(1/4)} \rho_{\rm i,kd}^{4/17}\tilde{r}^{9/17}_{\rm kd}\left(\frac{\sigma_{d}}{m_{\chi}}\right)^{4/17}H^{-4/17}(z)
& \qquad d{\rm -wave} 
  \end{cases}~,
\eeq
where the subscripts 3/2 (eq.~(\ref{A12})) and 9/4  (eq.~(\ref{A13})) indicate whether the maximum density profile terminates at a 3/2nd  profile or the 9/4th  profile.
   Using eq.~\eqref{eq:AnnihilationRate0} we next derive analytic forms for the annihilation rate at different redshifts  $\hat\Gamma_{\bullet}=\Gamma_{\bullet}(h(z))^x$ where $x$ is determined case-by-case, and is such that $\hat\Gamma_{\bullet}|_{z=0}=\Gamma_{\bullet}$.
   
   \subsubsection{Heavy Black Holes}
We begin by considering  heavy PBH in which case the density profile is determined by maximum density core followed by the 9/4th profile. Let us first give the $s$-wave case (where we use the notation $\hat{\Gamma}_{\bullet,~s}^{H}$ with the subscript $s$ indicating the  $s$-wave case and the superscript $H$ to indicate the heavy PBH scenario), for which
\beq
\hat{\Gamma}_{\bullet,~s}^{H}&=\frac{4\pi r_{\rm Sch}^{3}}{m_{\chi}^{2}}\left[\frac{m_{\chi}^{2}H^{2}(z)}{3\sigma_{s}}\left(\tilde{r}_{\rm core}\right)_{9/4}^{3}+\frac{128\pi}{\Gamma^{4}(1/4)}\rho_{\rm i,kd}^{2}\tilde{r}_{\rm kd}^{9/2}\left(\frac{2\sigma_{s}}{3}\right)\left[\left(\tilde{r}_{\rm core}\right)_{9/4}^{-3/2}-\tilde{r}_{\rm ta}^{-3/2}\right]\right],
\eeq
where $r_{\rm ta}$ is the radius of influence to be evaluated at the matter-radiation equilibrium. Simplifying one can obtain
\beq
\hat{\Gamma}_{\bullet,~s}^{H}&=\frac{4\pi r_{\rm Sch}^{3}}{m_{\chi}^{2}} \left[\frac{(128\pi)^{2/3}}{\Gamma^{8/3}(1/4)}m_{\chi}^{2/3}\rho_{\rm i,kd}^{4/3}\tilde{r}^{3}_{\rm kd}\sigma_{s}^{1/3}H^{2/3}(z) - \frac{256\pi}{3\Gamma^{4}(1/4)}\rho_{\rm i,kd}^{2}\tilde{r}_{\rm kd}^{9/2}\sigma_{s}\tilde{r}_{\rm ta}^{-3/2}\right].
\eeq
Note that the second term here is sub-leading and can typically be neglected.

Similarly, for the $p$-wave case one has
\beq
\hat{\Gamma}_{\bullet,~p}^{H}&=\frac{4\pi r_{\rm Sch}^{3}}{m_{\chi}^{2}}\left[\frac{m_{\chi}^{2}H^{2}(z)}{2\sigma_{p}}\left(\tilde{r}^{p}_{\rm core}\right)_{9/4}^{4} +\frac{128\pi}{5\Gamma^{4}(1/4)}\rho_{\rm i,kd}^{2}\tilde{r}_{\rm kd}^{9/2}\sigma_{p}\left[\left(\tilde{r}_{\rm core}\right)_{9/4}^{-5/2}-\tilde{r}_{\rm ta}^{-5/2}\right]\right]\\
& = \frac{4\pi r_{\rm Sch}^{3}}{m_{\chi}^{2}}\left[\frac{13(32\pi)^{8/13}}{10\Gamma^{32/13}(1/4)}m_{\chi}^{10/13}\sigma_{p}^{3/13}\rho_{\rm i,kd}^{16/13}\tilde{r}_{\rm kd}^{36/13}H^{10/13}(z)-\frac{128\pi}{5\Gamma^{4}(1/4)}\rho_{\rm i,kd}^{2}\tilde{r}_{\rm kd}^{9/2}\sigma_{p}\tilde{r}_{\rm ta}^{-5/2}\right].
\eeq
and the $d$-wave case
\beq
\hat{\Gamma}_{\bullet, ~d}^{H}&=\frac{4\pi r_{\rm Sch}^{3}}{m_{\chi}^{2}}\left[\frac{4m_{\chi}^{2}H^{2}(z)}{5\sigma_{d}}\left(\tilde{r}^{d}_{\rm core}\right)_{9/4}^{5} +\frac{64\pi}{7\Gamma^{4}(1/4)}\rho_{\rm i,kd}^{2}\tilde{r}_{\rm kd}^{9/2}\sigma_{d}\left[\left(\tilde{r}_{\rm core}\right)_{9/4}^{-7/2}-\tilde{r}_{\rm ta}^{-7/2}\right]\right],\\
& =  \frac{4\pi r_{\rm Sch}^{3}}{m_{\chi}^{2}}\left[\frac{68(8\pi)^{10/17}}{35\Gamma^{40/17}(1/4)}\rho_{\rm i,kd}^{20/17}\tilde{r}_{\rm kd}^{45/17}\sigma_{d}^{3/17}m_{\chi}^{14/17}H^{14/17}(z)-\frac{64\pi}{7\Gamma^{4}(1/4)}\rho_{\rm i,kd}^{2}\tilde{r}_{\rm kd}^{9/2}\sigma_{d}\tilde{r}_{\rm ta}^{-7/2}\right].
\eeq

\subsubsection{Light Black holes}
For lighter PBH the density profile is determined by the maximum density core transitioning into a 3/2 profile. In this 
case the rate for $s$-wave annihilating dark matter is given by
\beq
\hat{\Gamma}_{\bullet, ~s}^{L}&=\frac{4\pi r_{\rm Sch}^{3}}{m_{\chi}^{2}}\left[\frac{m_{\chi}^{2}H^{2}(z)}{3\sigma_{s}}\left(\tilde{r}_{\rm core}\right)_{3/2}^{3}+\frac{2\sigma_{s}}{\pi^{3}}\rho_{\rm i,kd}^{2}x_{\rm kd}^{3}\ln\left(\frac{\tilde{r}_{\rm ta}}{\left(\tilde{r}_{\rm core}\right)_{3/2}}\right)\right],\\
& = \frac{8r_{\rm Sch}^{3}}{3\pi^{2}}\rho_{\rm i,kd}^{2}x_{\rm kd}^{3}\frac{\sigma_{s}}{m_{\chi}^{2}}\left[1+\ln\left(\frac{m_{\chi}^{2}\tilde{r}_{\rm ta}^{3}H^{2}(z)}{2\rho_{\rm i,kd}^{2}x_{\rm kd}^{3}\sigma_{s}^{2}}\right)\right]\approx {\Gamma}_{\bullet, ~s}^{L}.
\eeq
An important observation is that the radiation due to the annihilations produced inside the core is independent of the redshift, while the contribution from the spike profile is only logarithmically sensitive to the expansion history of the universe, the redshift independent piece typically dominates.

Similarly for the $p$-wave case
\beq
\hat{\Gamma}^{L}_{\bullet, p} & = \frac{4\pi r_{\rm Sch}^{3}}{m_{\chi}^{2}}\left[\frac{m_{\chi}^{2}H^{2}(z)}{2\sigma_{p}}\left(\tilde{r}_{\rm core}\right)_{3/2}^{4}+\frac{\rho_{\rm i,kd}^{2}x_{\rm kd}^{3}\sigma_{p}}{\pi^{3}}\left[\left(\tilde{r}_{\rm core}\right)_{3/2}^{-1} - \tilde{r}_{\rm ta}^{-1}\right]\right],\\
& = \frac{4\pi r_{\rm Sch}^{3}}{m_{\chi}^{2}}\left[\frac{1}{2}\left(\frac{1}{2\pi^{3}}\right)^{4/5}m_{\chi}^{2/5}H^{2/5}\rho_{\rm i,kd}^{8/5}x_{\rm kd}^{12/5}\sigma_{p}^{3/5}\left\lbrace 1+4\left[1-\frac{\rho_{\rm i,kd}^{2/5}x_{\rm kd}^{3/5}}{(2\pi^{3})^{1/5}}
\left(\frac{\sigma_{p}}{m_{\chi}}\right)^{2/5}H^{-2/5}\tilde{r}_{\rm ta}^{-1}\right] \right\rbrace \right],
\eeq
and for the $d$-wave case
\beq
\hat{\Gamma}_{\bullet, ~d}^{L} & = \frac{4\pi r_{\rm Sch}^{3}}{m_{\chi}^{2}}\left[\frac{4m_{\chi}^{2}H^{2}(z)}{5\sigma_{d}}\left(\tilde{r}_{\rm core}\right)^{5}_{3/2}+\frac{\rho_{\rm i,kd}^{2}x_{\rm kd}^{3}\sigma_{d}}{4\pi^{3}}\left[\left(\tilde{r}_{\rm core}\right)^{-2}_{3/2} - \tilde{r}^{-2}_{\rm ta}\right]\right. 
\left. +\frac{64\pi}{7\Gamma^{4}(1/4)\rho_{\rm i,kd}^{2}\tilde{r}_{\rm kd}^{9/2}}\left(\tilde{r}_{B}^{\prime\prime -7/2} - \tilde{r}_{\rm ta}^{-7/2}\right)\right]\\
& = \frac{4\pi r_{\rm Sch}^{3}}{m_{\chi}^{2}}\left[\frac{1}{5}\left(\frac{1}{2\pi^{15}}\right)^{\frac{1}{7}}m_{\chi}^{4/7}H^{4/7}\rho_{\rm i,kd}^{10/7}x_{\rm kd}^{15/7}\sigma_{d}^{3/7}\left\lbrace 1+\frac{5}{2}\left[ 1-\left(\frac{1}{8\pi^{3}}\right)^{2/7}\rho_{\rm i,kd}^{4/7}x_{\rm kd}^{6/7}\left(\frac{\sigma_{d}}{m_{\chi}}\right)^{4/7}\frac{H^{-4/7}}{\tilde{r}_{\rm ta}^{2}}\right]\right\rbrace\right].
\eeq

\subsubsection{Intermediate Mass Black holes}
Finally, we consider the case of an intermediate mass PBH (as defined in Section \ref{2.2}) in which the maximum density core transitions into a 3/2 density profile, which subsequently transitions into a 9/4 profile.
For the $s$-wave case the annihilation rate is given by
\beq
\hat{\Gamma}_{\bullet, ~s}^{I}&=\frac{4\pi r_{\rm Sch}^{3}}{m_{\chi}^{2}}\Big[\frac{m_{\chi}^{2}H^{2}(z)}{3\sigma_{s}}\left(\tilde{r}_{\rm core}\right)_{3/2}^{3}+\frac{2\sigma_{s}}{\pi^{3}}\rho_{\rm i,kd}^{2}x_{\rm kd}^{3}\ln\left(\frac{\tilde{r}_{B}^{\prime\prime}}{\left(\tilde{r}_{\rm core}\right)_{3/2}}\right)\\
&\hspace{4cm}+\frac{128\pi}{\Gamma^{4}(1/4)}\rho_{\rm i,kd}^{2}\tilde{r}_{\rm kd}^{9/2}\left(\frac{2\sigma_{s}}{3}\right)\left[\left(r_{B}^{\prime\prime}\right)^{-3/2}-\tilde{r}_{\rm ta}^{-3/2}\right]\Big],\\[8pt]
& = \frac{8r_{\rm Sch}^{3}}{3\pi^{2}}\rho_{\rm i,kd}^{2}x_{\rm kd}^{3}\frac{\sigma_{s}}{m_{\chi}^{2}}\Big[1+\ln\left(\frac{2048\pi^{11}}{\Gamma^{8}(1/4)}\rho_{\rm i,kd}^{-2}\tilde{r}_{\rm kd}^{9}x_{\rm kd}^{-9}\sigma_{s}^{-2}m_{\chi}^{2}H^{2}\right) \\
&\hspace{4cm} + \frac{128\pi^{4}}{\Gamma^{4}(1/4)}\tilde{r}_{\rm kd}^{9/2}x_{\rm kd}^{-3}\left[\frac{\Gamma^{4}(1/4)}{64\pi^{4}}\tilde{r}_{\rm kd}^{-9/2}x_{\rm kd}^{3}-\tilde{r}_{\rm ta}^{-3/2}\right]\Big].
\eeq
For the $p$-wave case the annihilation rate is given by
\beq
\hat{\Gamma}_{\bullet, ~p}^{I} & = \frac{4\pi r_{\rm Sch}^{3}}{m_{\chi}^{2}}\left[\frac{2m_{\chi}^{2}H^{2}(z)}{\sigma_{p}}\frac{\left(\tilde{r}_{\rm core}\right)^{4}_{3/2}}{4}+\frac{\rho_{\rm i,kd}^{2}x_{\rm kd}^{3}\sigma_{p}}{\pi^{3}}\left[\left(\tilde{r}_{\rm core}\right)^{-1}_{3/2} - \tilde{r}^{\prime\prime -1}_{B}\right]\right.  \left. +\frac{128\pi}{5\Gamma^{4}(1/4)}\rho_{\rm i,kd}^{2}\tilde{r}^{9/2}_{\rm kd}\sigma_{p}\left[\tilde{r}_{B}^{\prime\prime -5/2} - \tilde{r}_{\rm ta}^{-5/2}\right]\right]\\[8pt]
& = \frac{4\pi r_{\rm Sch}^{3}}{m_{\chi}^{2}}\left[\frac{1}{2}\left(\frac{1}{2\pi^{3}}\right)^{4/5}\rho_{\rm i,kd}^{8/5}x_{\rm kd}^{12/5}\sigma_{p}^{3/5}m_{\chi}^{2/5}H^{2/5}(z)\left(5 -4\left(2\pi^{3}\right)^{-1/5}\rho_{\rm i,kd}^{2/5}x_{\rm kd}^{3/5}\left(\frac{\sigma_{p}}{m_{\chi}}\right)^{2/5}H^{-2/5}(z)\tilde{r}_{B}^{\prime\prime -1} \right)\right. \\
&\hspace{2cm} \left. +\frac{128\pi}{5\Gamma^{4}(1/4)}\rho_{\rm i,kd}^{2}\tilde{r}_{\rm kd}^{9/2}\sigma_{p}\left[\tilde{r}_{B}^{\prime\prime -5/2} - \tilde{r}_{\rm ta}^{-5/2}\right]\right].
\eeq
Finally, for the $d$-wave case one has
\beq
\hat{\Gamma}_{\bullet, ~d}^{I} & = \frac{4\pi r_{\rm Sch}^{3}}{m_{\chi}^{2}}\left[\frac{4m_{\chi}^{2}H^{2}(z)}{5\sigma_{d}}\left(\tilde{r}_{\rm core}\right)^{5}_{3/2}+\frac{\rho_{\rm i,kd}^{2}x_{\rm kd}^{3}\sigma_{d}}{4\pi^{3}}\left[\left(\tilde{r}_{\rm core}\right)^{-2}_{3/2} - \tilde{r}^{\prime\prime -2}_{B}\right] 
 +\frac{64\pi}{7\Gamma^{4}(1/4)\rho_{\rm i,kd}^{2}\tilde{r}_{\rm kd}^{9/2}}\left(\tilde{r}_{B}^{\prime\prime -7/2} - \tilde{r}_{\rm ta}^{-7/2}\right)\right]\\[8pt]
& = \frac{4\pi r_{\rm Sch}^{3}}{m_{\chi}^{2}}\left[\frac{1}{5}\left(\frac{1}{2\pi^{15}}\right)^{1/7}m_{\chi}^{4/7}H^{4/7}(z)\rho_{\rm i,kd}^{10/7}x_{\rm kd}^{15/7}\sigma_{d}^{3/7}\left[ \frac{7}{2}-\frac{5}{2}\left(\frac{1}{8\pi^{3}}\right)^{2/7}\rho_{\rm i,kd}^{4/7}x_{\rm kd}^{6/7}\left(\frac{\sigma_{d}}{m_{\chi}}\right)^{4/7}\frac{H^{-4/7}(z)}{\tilde{r}_{B}^{\prime\prime 2}}\right]\right.\\
&\hspace{2cm}  \left. + \frac{64\pi}{7\Gamma^{4}(1/4)}\sigma_{d}\rho_{\rm i,kd}^{2}\tilde{r}_{\rm kd}^{9/2}\left(\tilde{r}_{B}^{\prime\prime -7/2}-\tilde{r}_{\rm ta}^{-7/2}\right)\right].
\eeq

\begin{figure}[t!]
\centerline{
 \includegraphics[scale=0.44]{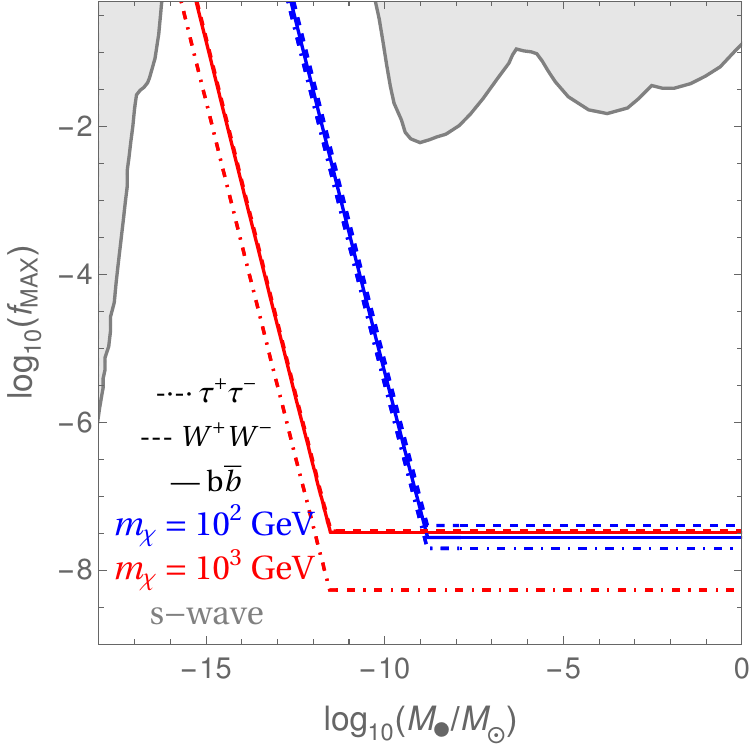}
 \includegraphics[scale=0.44]{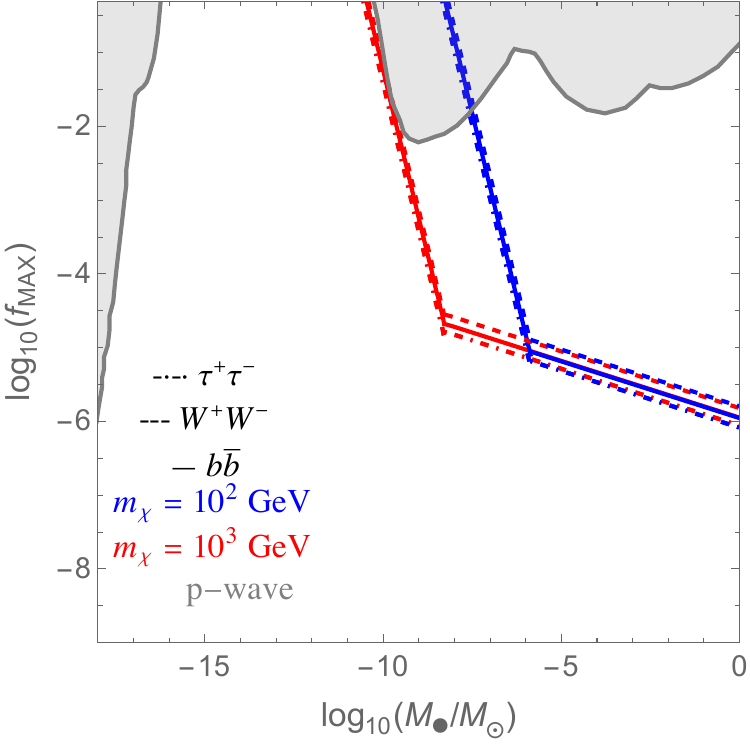}
 \includegraphics[scale=0.44]{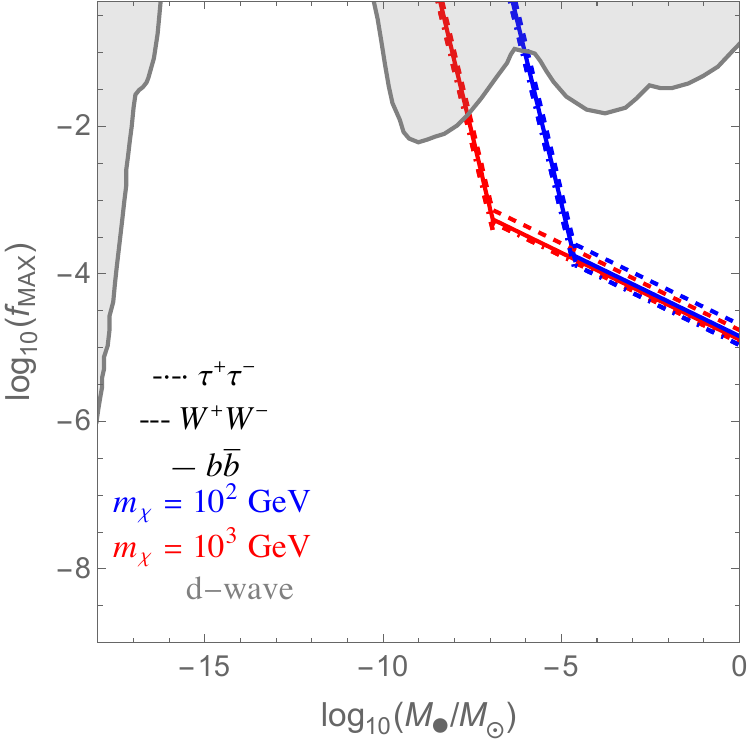}}
\caption{The solid curves are identical to Figure \ref{Fig:sWave-fboundMBH-Galactic} and show the extragalactic $\gamma$-ray bounds assuming 100\% of the dark matter annihilation primaries are $b\overline{b}$ pairs . The other curves also show the extragalactic $\gamma$-ray bounds, but where we vary our assumptions regarding the annihilation channel with 100\% $W^+W^-$ pairs shown dashed, and 100\% $\tau^+\tau^-$ pairs shown dot-dashed.
\vspace{-3mm}
  \label{Wtau}}
\end{figure}

\newpage
   \subsection{Annihilation Channel Dependance of Gamma-Ray Bounds} \label{A5}

To derive the bounds on dark matter annihilations one must define the photon spectrum due to annihilations ${\rm d}N_{\gamma}/{\rm d}E$. The photon spectrum depends on the details of the model, and in the plots presented in Figures \ref{Fig:sWave-fboundMBH-Galactic} \& \ref{Fig:svp2}, we have made the simplistic (although not unreasonable) assumption that 100\% of the primary annihilation products are $b\overline{b}$ pairs. Such a scenario  naturally occurs if the dark matter annihilates via a scalar mediator that mixes with the Standard Model Higgs and thus inherits the hierarchical structure of the Standard Model Yukawa couplings, for instance, see e.g.~\cite{Martin:2014sxa}.

It is immediately clear, however, that one can construct dark matter scenarios in which the annihilation products are not entirely $b\overline{b}$ pairs, or indeed, some other state provides dominant channel. Without a specific dark matter model in mind it is not feasible to consider all possible annihilation channels, however, we find it prudent to consider some alternative channels (although to our knowledge prior studies have not investigated this issue). 

In Figure  \ref{Wtau} we present an analogous plot to Figure \ref{Fig:sWave-fboundMBH-Galactic}, but where we show only the extragalactic $\gamma$-ray bounds and we vary our assumptions regarding the annihilation channel. The 100\% $b\overline{b}$ pairs shown as solid lines in Figure \ref{Wtau} (matching Figure \ref{Fig:sWave-fboundMBH-Galactic}), 100\% $W^+W^-$ pairs shown dashed, and 100\% $\tau^+\tau^-$ pairs shown dot-dashed. For each different annihilation channel we use the corresponding photon spectrum ${\rm d}N_{\gamma}/{\rm d}E$, taking these distributions from  \cite{Cembranos:2010dm}. Changes between these three annihilation channels typically result in $\mathcal{O}(1)$ differences in the limits and does not significantly impact the broad conclusions of our analysis.

\begin{figure}[b!]
\centerline{
 \includegraphics[width=0.46 \textwidth]{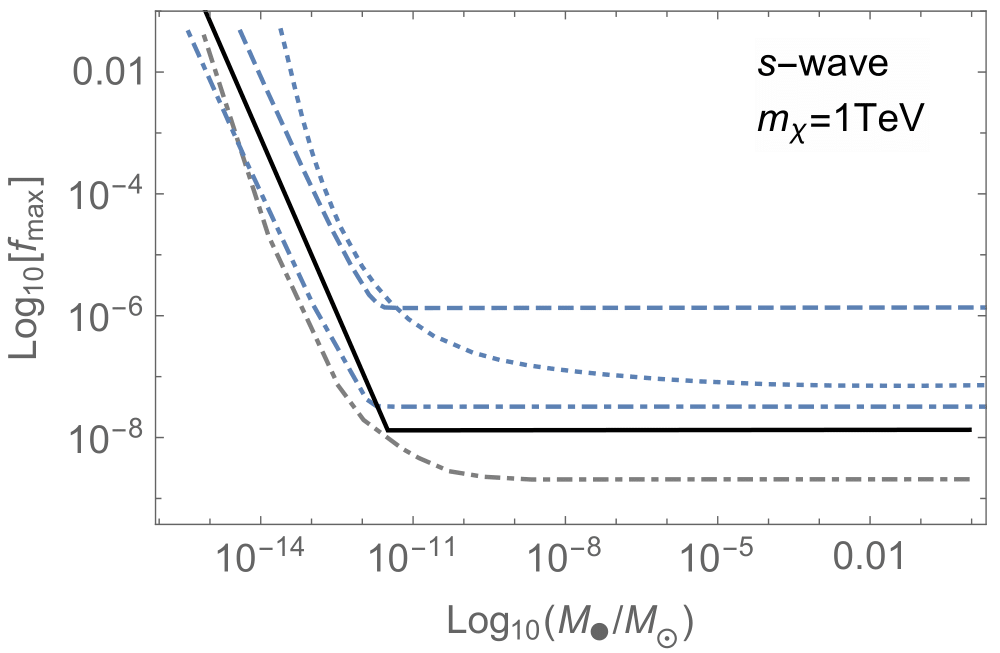}
 \hspace{5mm}
 \includegraphics[width=0.46 \textwidth]{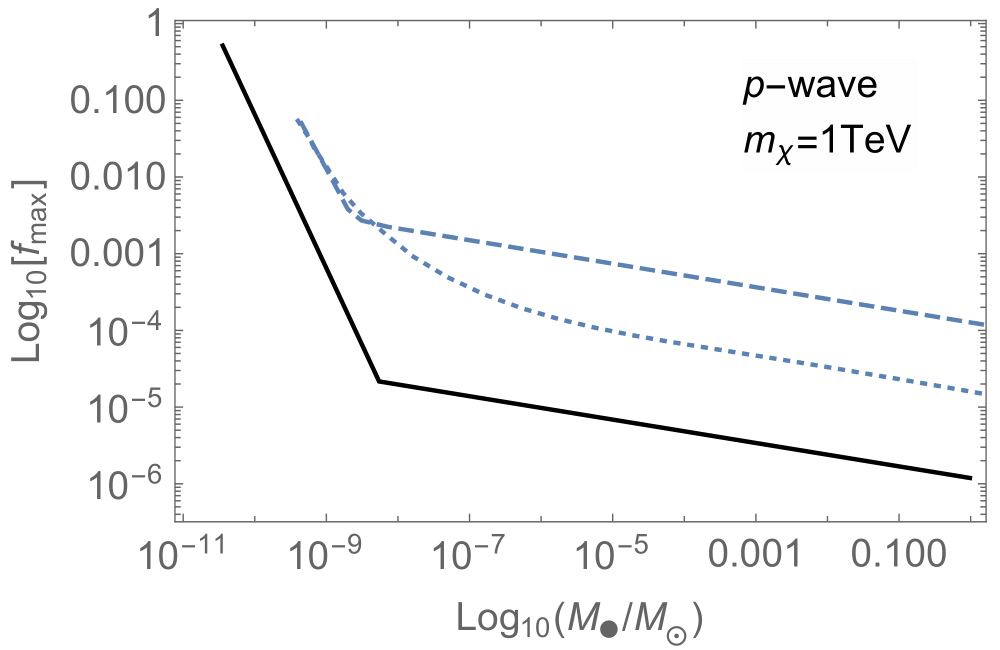}}
\centerline{
  \includegraphics[width=0.46 \textwidth]{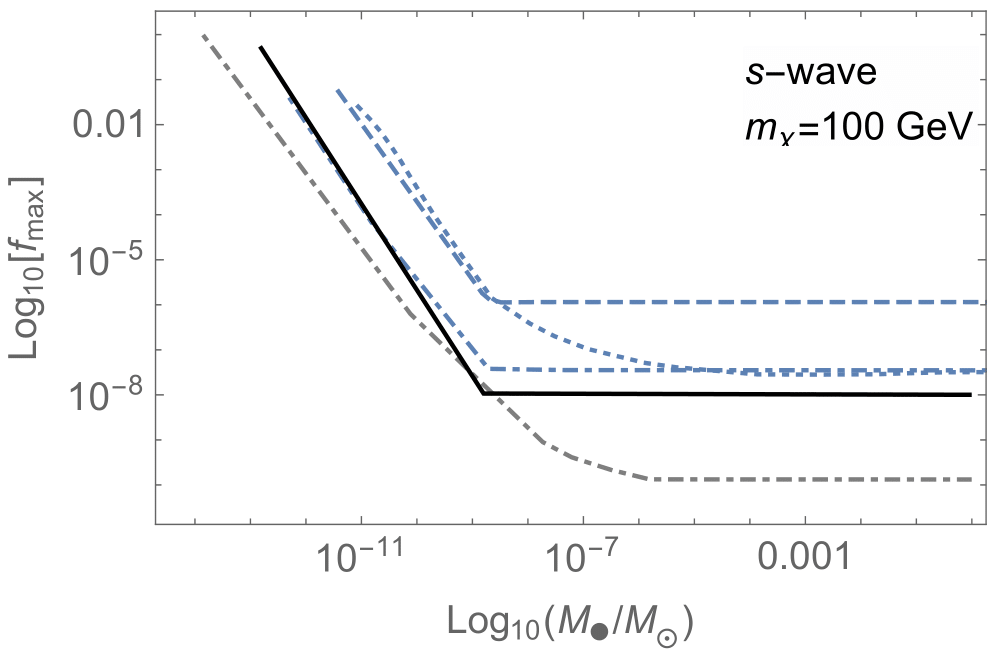}
 \hspace{5mm}
 \includegraphics[width=0.46 \textwidth]{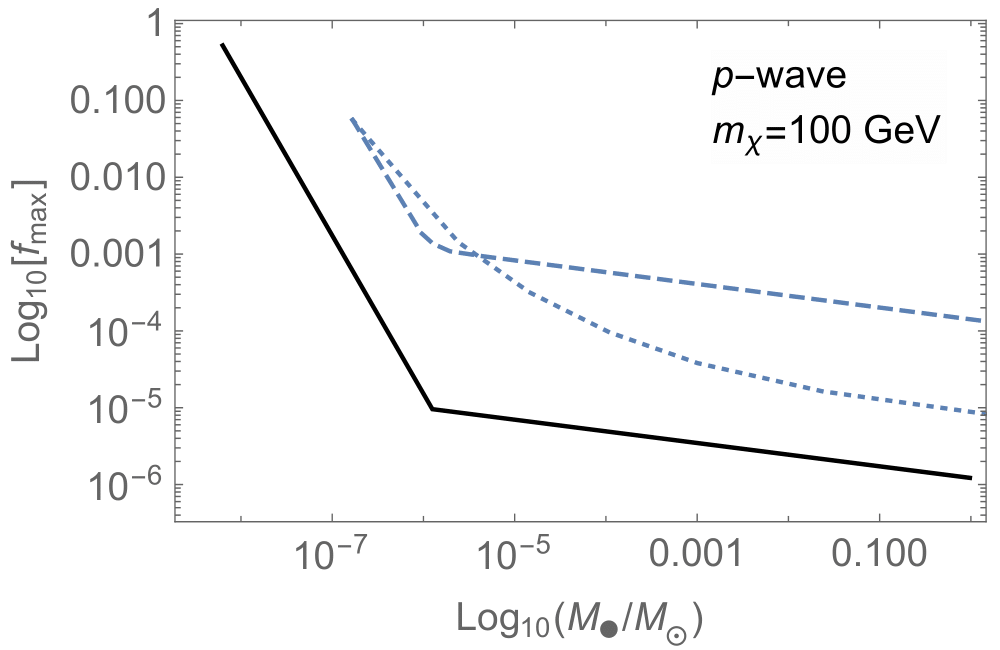}}
\vspace{-3mm}
\caption{A comparison to alternative constraints, presented analogously to Figure \ref{Fig:sWave-fboundMBH-Galactic}. 
The upper bound on the fractional abundance of PBH $f_{\rm PBH}\lesssim f_{\rm MAX}$ as the PBH mass $M_\bullet$ is varied for different dark matter scenarios and different experimental limits. Our limits from the extragalactic $\gamma$-ray background are shown as black, solid curves. The analogous from extragalactic $\gamma$-ray bound derived by \cite{Gines:2022qzy} is shown as blue dashed. We also show  the limits from \cite{Gines:2022qzy} for CMB constraints (blue dotted), and from recasting bounds from decaying dark matter (blue dot-dashed). We overlay (grey dashed) the $s$-wave extragalactic $\gamma$-ray bound derived in \cite{Carr:2020mqm}, but importantly highlight that this analysis propagates an error in the literature relating to application of the $\gamma$-ray constraint and is not reliable.
\label{Fig:new2}}
\end{figure}

   \subsection{Comparison to other Analyses}
   \label{A7}
   
We have extracted limits from the extragalactic $\gamma$-ray flux from Fermi-LAT  \cite{Fermi-LAT:2015qzw}. Other groups have examined this data, as well as alternative constrains. In Figure \ref{Fig:new2} we take our results of Figure \ref{Fig:sWave-fboundMBH-Galactic} and overlay these with the results of other groups to provide a comparison. 

\subsubsection{Extragalactic $\gamma$-ray flux}

We now compare our results (solid black curves in Figure \ref{Fig:new2}) to  analogous analyses in \cite{Gines:2022qzy} (blue dashed) and \cite{Carr:2020mqm} (grey dashed). Comparing to  \cite{Gines:2022qzy} (based on the same underlying data \cite{Fermi-LAT:2015qzw}), our limits are a factor of 100 stronger. Notably, our dark matter halo profiles have inner density regions with $\mathcal{O}(1)$ larger inner profile radii.  Additionally, while we have fixed $x_{\rm KD}$ in our analyses, in \cite{Gines:2022qzy} the authors apply some scaling relationship. Resultantly, the flux in \cite{Gines:2022qzy} and our own analyses differ by roughly a factor of $10$. The other factor of $10$ difference coming from the application of indirect detection bounds.  Overall, we consider our results to be largely compatible, and these limits should be taken as characteristic since the exact model details can lead to modest changes (cf.~the mixed $s$-wave/$p$-wave case in Section \ref{S6}). Further, neither our analysis or the method of \cite{Gines:2022qzy} is especially sophisticated in applying the indirect detection constraints and certainly a more careful analysis might be conducted (see for instance discussions in \cite{Cirelli:2012ut}), although deviations in limits are anticipated to be modest.

We note that the $s$-wave extragalactic $\gamma$-ray bound derived in \cite{Carr:2020mqm} (grey dashed), while stronger, is in error. The work propagates an error in the earlier literature relating to application of the $\gamma$-ray constraint and is not reliable. It is presented here for comparison purposes only. Specifically, this issue comes from the specific criteria for applying the indirect detection limits when calculating $f_{\rm MAX}$. Other aspects of \cite{Carr:2020mqm} are correct and insightful.

\subsubsection{Other approaches}

It was highlighted in  \cite{Boucenna:2017ghj} that one might obtain a constraint on annihilations in PBH dark matter halos through matching the limits on the decaying dark matter, with decay rate $\Gamma_{\rm DDM}$. Specifically, one may make the following identification  with the annihilation rate in the PBH halo $\Gamma_{\bullet}$
\beq
\frac{f_{\rm PBH}(1-f_{\rm PBH})^2\Gamma_{\bullet}}{M_\bullet}=  \frac{\Gamma_{\rm DDM}}{m_{\chi}}~.
\eeq
In this manner one can potentially recast decaying dark matter limits to place bounds on PBH with dark matter halos (this is the specific method used in the recent papers of \cite{Kadota:2021jhg,Gines:2022qzy} to set a bound on $p$-wave annihilations around PBH).
It was argued that for the dark matter mass range, $m_{\chi} = 1-10^{4}~{\rm GeV}$, by inspection of  \cite{Ando:2015qda}, a reasonable reference value for the limit on the lifetime of dark matter is $\tau_{\rm DM}=\Gamma_{\rm DDM}^{-1}\gtrsim 10^{28}~{\rm s}$ based on \cite{Fermi-LAT:2014ryh}. From this limit one can extract a constraint on $f_{\rm MAX}$ for a given $M_\bullet$ and dark matter model. The results from \cite{Gines:2022qzy} using these recast limits are shown as dot-dashed curves. We also show the CMB constraints derived in \cite{Gines:2022qzy}, shown as blue dotted curves. It is also worth highlighting that these studies use a simplified treatment of the halo redshift dependance (e.g.~using only eq.~(\ref{GB})).

We  strongly favour using Fermi-LAT data directly, over recast decaying dark matter limits, as we believe that it is much more robust to match to a primary experimental data source rather than try to interpolate from a very different analysis using broad order of magnitude assumptions regarding critical values (in this case $m_{\rm DM}$). Notably, the limits from these recast decaying dark matter limits and CMB constraints are comparable  to the limits from extragalactic $\gamma$-rays derived here and thus are complementary.

\end{document}